\documentclass[fleqn,usenatbib]{mnras}

\usepackage[T1]{fontenc}

\DeclareRobustCommand{\VAN}[3]{#2}
\let\VANthebibliography\thebibliography
\def\thebibliography{\DeclareRobustCommand{\VAN}[3]{##3}\VANthebibliography}

\usepackage{graphicx}	% Including figure files
\usepackage{amsmath}	% Advanced maths commands
\usepackage{amssymb}	% Extra maths symbols
\usepackage{txfonts}
\usepackage{epstopdf}
\usepackage{comment}
\usepackage{xcolor,colortbl}
\usepackage{multirow}
\usepackage{lscape} %texstudio
\usepackage{threeparttable} %texstudio
\title[Disk-wind regulated accretion in 4U 1543$-$47]{Wideband Study of the Brightest Black Hole X-ray Binary 4U 1543$-$47 in the 2021 Outburst: Signature of Disk-Wind Regulated Accretion}

\author[Prabhakar et al.]{
Geethu Prabhakar$^{1}$\thanks{geethuprabhakar.17@res.iist.ac.in},
Samir Mandal$^{1}$\thanks{samir@iist.ac.in},
Bhuvana G. R.$^{2}$,
and Anuj Nandi$^{3}$
\\
$^{1}$ Department of Earth and Space Sciences, Indian Institute of Space Science and Technology (IIST), Trivandrum - 695547, India\\
$^{2}$ Department of Physics, Dayananda Sagar University, Bengaluru - 560068, India\\
$^{3}$ Space Astronomy Group, ISITE Campus, U R Rao Satellite Centre, Bengaluru - 560037, India
}

\date{Accepted XXX. Received YYY; in original form ZZZ}

\pubyear{2022}

\begin{document}
\label{firstpage}
\pagerange{\pageref{firstpage}--\pageref{lastpage}}
\maketitle

% Abstract of the paper
\begin{abstract}
A comprehensive wideband spectral analysis of the brightest black hole X-ray binary 4U $1543-47$ during its 2021 outburst is carried out for the first time using \textit{NICER, NuSTAR,} and \textit{AstroSat} observations by phenomenological and reflection modelling. The source attains a super-Eddington peak luminosity and remains in the soft state, with a small fraction ($< 3\%$) of the inverse-Comptonized photons. The spectral modelling reveals a steep photon index ($\Gamma \sim 2-2.6$) and relatively high inner disk temperature (T$_{in}\sim 0.9-1.27$ keV). 
The line-of-sight column density varies between ($0.45-0.54$)$\times10^{22}$ cm$^{-2}$. Reflection modelling using the RELXILL model suggests that 4U $1543-47$ is a low-inclination system ($\theta \sim 32^\circ - 40^\circ$). 
The accretion disk is highly ionized (log $\xi$ > 3) and has super solar abundance (3.6$-$10 $A_{Fe,\odot}$) over the entire period of study. We detected a prominent dynamic absorption feature between $\sim 8-11$ keV in the spectra throughout the outburst. This detection is the first of its kind for X-ray binaries.
We infer that the absorption of the primary X-ray photons by the highly ionized, fast-moving disk-winds can produce the observed absorption feature. The phenomenological spectral modelling also shows the presence of a neutral absorption feature $\sim 7.1 - 7.4$ keV, and both ionized and neutral absorption components follow each other with a delay of a typical viscous timescale of $10-15$ days. 
%Finally, we discuss the possible implication of our findings.
\end{abstract}

\begin{keywords}
accretion, accretion disc - black hole physics - X-rays: binaries - stars: individual: 4U $1543-47$
\end{keywords}

%%%%%%%%%%%%%%%%%%%%%%%%%%%%%%%%%%%%%%%%%%%%%%%%%%

%%%%%%%%%%%%%%%%% BODY OF PAPER %%%%%%%%%%%%%%%%%%

\section{Introduction}
X-ray spectroscopy of black hole X-ray binaries (BH-XRBs) holds the key to unveil the geometry of the system and the dynamics of the accretion process.
The spectrum of BH-XRBs mainly consists of a hard powerlaw and a soft thermal component. The soft component, which is a multi-temperature blackbody,
is assumed to be originated from an optically thick, geometrically thin accretion disk \citep{1973A&A....24..337S}. The hard powerlaw component is generally believed to be emitted from an optically thin, hot electron cloud called `corona' by the Comptonization of the soft disk photons \citep{1980A&A....86..121S,1985A&A...143..374S,1994MNRAS.269L..55Z,1995ApJ...455..623C,1998PhST...77...57P,2006ApJ...642L..49C,2015ApJ...807..108I,2018A&A...614A..79P}.
The relative strength of these components leads to different states in outbursting BH-XRBs. In the Low/Hard State (LHS), the non-thermal component dominates and in the High/Soft State (HSS), the disk emission dominates. 
There are short-lived intermediate states also, namely, the Hard Intermediate State (HIMS) and Soft Intermediate State (SIMS), lying between LHS and HSS \citep{2001ApJS..132..377H,2005Ap&SS.300..107H,2006ARA&A..44...49R,2012A&A...542A..56N,2019MNRAS.487..928S,2019MNRAS.486.2705A,2021MNRAS.501.5457B,2022MNRAS.514.6102P}. A typical outburst starts with the LHS and proceeds through intermediate states to HSS then back to LHS again and finally reach quiescence. However, it does not always have to go through all the states mentioned above \citep{2015ApJ...803...59D,2016MNRAS.460.4403R,2019ApJ...885...48G,2020MNRAS.497.1197B,2021MNRAS.508.2447B,2022MNRAS.514.6102P}.

The advent of high resolution spectroscopy reveals the presence of reflection features in the spectra of many BH-XRBs.
Irrespective of the geometry of the corona, it is believed that the photons upscatterd by the corona, the primary photons interact with the disc material and a part of which produces the reflection features \citep{1974A&A....31..249B}.
The reprocessed X-ray spectrum consists of fluorescent line emission from various elements, a soft thermal continuum and a Compton hump peaked at $\sim 20-30$ keV.
The most prominent feature among the fluorescent emission lines is the iron K-edge at $\sim 7.1$ keV \citep{1998ApJ...492..782U} and K$\alpha$ line at $\sim 6-7$ keV  \citep{1986MNRAS.218..129W,2002ApJ...576..391B,2005ApJ...623L.121D}. This is because the fluorescence yield increases with the atomic number \citep{1952burhop}.
For a distant observer, these reflection features appeared to be diluted/broadened and distorted (asymmetric) due to relativistic effects of the strong gravity region in the close vicinity of the BH \citep{1989MNRAS.238..729F,2000PASP..112.1145F}.
Spectral modelling using relativistic reflection models can address the effect of blurring of the spectral features and helps to probe the physics of the strong gravity region at the inner disk. The accretion disk characteristics, such as the ionization of the disk material, the iron abundance, inclination of the system, spin of the BH etc., can also be obtained from reflection modelling. The line broadening can also be due to Comptonization in a highly ionized, optically thick cloud, and the resultant feature is broad and symmetric.  However, this mechanism is important for high inclination systems only \citep{2001MNRAS.328..501P}.

The Fe$-$K band ($5-8$ keV) is the energy range where most of the emission/absorption features appear. 
The first observational evidence of the Fe$-$K absorption lines was provided by \citet{1998ApJ...492..782U} with \textit{ASCA} in the spectra of galactic superluminal BH source GRO J$1655-40$. \citet{2000ApJ...539..413K} and \citet{2002ApJ...567.1102L} also  detected similar features in the superluminal jet source GRS $1915+105$ with \textit{ASCA}. Later, it is revealed that the absorption features are very common in the spectra of BH-XRBs \citep{2013ApJ...779...26S,2014ApJ...784L...2K,2018ApJ...865...18X}. 
Photon interaction with neutral and static material produces sharp fluorescent lines at their corresponding transition energy levels. In case of ionized absorbers, there would be an increase in the transition line energy compared to their neutral ones.
 The absorption lines from highly ionized ions give an insight into the highly ionized plasma around the compact object.

The process of accretion in XRBs is usually accompanied with outflows and/or jets \citep{1999MNRAS.304..865F,2004MNRAS.355.1105F,2010MNRAS.406.1425F,2012ApJ...759L...6M,2013ApJ...775L..45M,2014AdSpR..54.1678R,2016MNRAS.460.4403R}. 
The persistent jets are present in the LHS of the system, and it gets turned off in HSS. Accretion disk-wind is generally observed in the disk-dominated HSS, though it can exist in other spectral states as well \citep{2002ApJ...567.1102L,2008ApJ...680.1359M,2009Natur.458..481N,2013AdSpR..52..732N}. The disk-winds carry a sufficient amount of matter which suppresses the launch of radio jets \citep{2009Natur.458..481N} in HSS. The disk-wind can also be highly ionized and their presence can be inferred by the blue-shifted absorption features in the X-ray spectrum \citep{1997xisc.conf..427E,1997AIPC..410..922K}. 
In general, it seems that the absorption lines are absent in the LHS, which is still a matter of debate. \citet{2009Natur.458..481N} suggests that the wind gets photoionized completely in LHS, and the medium becomes transparent; this could be a possible reason for the absence of absorption lines in the spectra.
%In LHS, the wind gets photoionized completely, and the medium becomes transparent, leading to the absence of absorption lines in the spectra \citep{2009Natur.458..481N}. 
Usually, the disk-winds are observed in high inclination systems \citep{2012MNRAS.422L..11P}. Such systems may show intensity `dips' in their X-ray spectra, for e.g.,  GRS $1915+105$, 4U $1630-47$, H $1743-322$, MAXI J$1305-704$, GRO J$1655-40$ \citep{1997MNRAS.287..622L,1998ApJ...494..753K,2013ApJ...779...26S}. The dips are believed to be caused by obscuring material associated with the accretion disk \citep{1987A&A...178..137F} and are visible for highly inclined systems with inclination angle $60^\circ \lesssim \theta \lesssim 80^\circ$ \citep{1987A&A...178..137F}.
The disk-winds play a major role in regulating the accretion scenario of BH-XRBs. For example, \citet{2016Natur.534...75M} showed how winds control the violent outburst of V404 Cygni by diminishing a significant fraction of the outer disk. Disk-wind studies in BH-XRBs can provide great insights into the physical mechanisms involved in the accretion process.

%---------------- 4U 1543−47 ---------------
4U $1543-47$ is a BH-XRB, discovered by \textit{Uhuru} satellite in 1971 \citep{1972ApJ...174L..53M}. Since the discovery, it has undergone five outbursts; the first four are in an interval of $\sim$ 10 years, in  1984 \citep{1984PASJ...36..799K}, 1992 \citep{1992IAUC.5504....1H} and 2002 \citep{2004ApJ...610..378P}. After a gap of 19 years, the fifth outburst happened in 2021 \citep{2021ATel14701....1N}, which marks the source as the brightest BH-XRB with a peak X-ray intensity of 11 Crab in $2-4$ keV with \textit{MAXI/GSC} \citep{2021ATel14708....1N}. The 2002 outburst was also brighter (4 Crab in $2-12$ keV), while the previous three outbursts have comparable intensities \citep{2004ApJ...610..378P}. 
Its optical counterpart, IL Lupi, was discovered by \citet{1983Msngr..34...21P}.
The central engine is a dynamically confirmed BH with a mass of $9.4 \pm 1.0$ M$_\odot$, and the companion is an A2V star of mass $2.45 \pm 0.15$ M$_\odot$ \citep{2006MNRAS.371.1334R}.
It is located at $RA=15^h 47^m 8^s.27$, $Dec=-47^\circ 40^{'} 10^{''}.8$ (J2000) \citep{2004ApJ...610..378P} at a distance of $7.5 \pm 0.5$ kpc \citep{2004MNRAS.354..355J}. \citet{2003IAUS..212..365O} estimated the orbital inclination of the system as $20.7^\circ \pm 1.5^\circ$.
There were multiple attempts to estimate the spin ($a_\ast$, dimensionless spin parameter) of the BH in 4U $1543-47$ using \textit{RXTE} observations of the 2002 outburst.
\citet{2006ApJ...636L.113S} estimated a spin of $\sim 0.75-0.85$ using continuum-fitting of \textit{RXTE} data. 
\citet{2009ApJ...697..900M} and \citet{2014ApJ...793L..33M} estimated the spin value as $0.3\pm 0.1$ and $0.43^{+0.22}_{-0.31}$ respectively using relativistic disk reflection and disk continuum modelling. These three estimations are based on a BH mass of $9.4 \pm 1.0$ M$_\odot$ and 
a distance of $7.5 \pm 0.5$ kpc. \citet{2006ApJ...636L.113S} and \citet{2014ApJ...793L..33M} used the binary inclination ($\theta$) of $20.7 \pm 1.5$ degree, while \citet{2009ApJ...697..900M} used a $\theta$ of $32_{-4}^{+3}$ degree for the spin estimation.
\citet{2020MNRAS.493.4409D} reported a spin of  $0.67^{+0.15}_{-0.08}$ and $\theta$ of $36.3_{-3.4}^{+5.3}$ degree by reflection modelling of \textit{RXTE} data using the model RELXILL. 

The Giant Metrewave Radio Telescope (\textit{GMRT}) detected radio flares from the source in 2002 outburst \citep{2004ApJ...610..378P}. Multiple flaring occasions are reported at different phases of the outburst. \citet{2020MNRAS.495..182R} reported the presence of a compact jet in the SIMS of the 2002 outburst of 4U $1543-47$ using multiwavelength observations (X-ray, optical, IR, and radio). Since the system has a low inclination, the jet angle and axis of rotation may coincide. 
\citet{2020MNRAS.495..182R} tested the chances of jet contribution to the luminosity of the system and renounced that possibility.

Until now, there is no study in literature based on the 2021 outburst of 4U $1543-47$. We aim for a detailed analysis of the wideband spectral characteristics of the 2021 outburst using three different instrument data from \textit{NICER} (\textit{Neutron
star Interior Composition ExploreR}), \textit{NuSTAR} (\textit{Nuclear Spectroscopic Telescope Array}) and \textit{AstroSat} during outburst decay. The evolution of spectral parameters is investigated using  phenomenological and reflection modelling.
Even though the reflection modelling of \textit{RXTE} data of 2002 outburst \citep{2009ApJ...697..900M,2014ApJ...793L..33M,2020MNRAS.493.4409D} unveil the fundamental quantities of the system like $a_\ast$ and $\theta$, data from much better spectral resolution instruments like \textit{NuSTAR} \citep{2013ApJ...770..103H} are highly promising. It can also provide outburst specific quantities like the iron abundance and ionization of the accretion disk. We report the presence of strong and dynamic absorption features in the 2021 outburst spectra, which has not been observed in any previous outbursts of 4U $1543-47$. We examine these features quantitatively using phenomenological modelling of \textit{NuSTAR} data.

This paper is structured as follows: The observations and the data reduction procedure are discussed in \S\ref{sec:redn}. The evolution of the outburst lightcurve and hardness ratio are examined in \S\ref{sec:lc-hr}.
The spectral modelling and parameter evolution are presented in \S\ref{sec:spec-result}. Phenomenological and reflection modelling of different epochs are discussed in \S\ref{sec:pheno-modelling} and \S\ref{sec:reflec-modelling}, respectively.
The detailed study of the absorption features in the spectra of 4U $1543-47$ is carried out in \S\ref{sec:absorption}. We discussed our overall findings in \S\ref{sec:discuss}. Finally, we summarise the results in \S\ref{sec:concl} and then conclude.

\begin{table*}
	\centering
	\caption{The list of observations of the source 4U $1543-47$ considered for the study. There are 16 epochs consisting of ten \textit{NuSTAR} and six \textit{AstroSat} observations. Seven \textit{NuSTAR} epochs have simultaneous \textit{NICER} coverage also.}
	\label{tab:allobs}
	\begin{tabular}{cccccc}  
		\hline
		\multirow{2}{*}{Epoch} &  \multicolumn{3}{c}{Obs. ID (MJD)} & \multirow{2}{*}{Remarks}\\
		\cline{2-4} 
		  & \textit{NuSTAR} & \textit{NICER} & \textit{AstroSat} &  \\

		\hline
		&&&\\
		1 & 80702317002 (59382.42) & 4655060101 (59382.44) & &\\ %a
		2 & 80702317004	(59389.47) & 4655060201	(59389.47) & &\\
		3 &  &  & T04\_018T01\_9000004494 (59396.04) & Offset\\
		4 & 80702317006 (59396.18) & 4655060301	(59396.19) & & \\ %b
		5 & 80702317008	(59403.02) & 4655060401 (59403.04) & &\\ %c
		6 &  &  & T04\_021T01\_9000004526 (59405.36) & Offset\\
		7 &  &  & T04\_030T01\_9000004588 (59421.19) & Pointed\\
		8 & 90702326002	(59421.67) & & &\\
		9 & 90702326004	(59428.18) & 4202230143	(59428.13) & &\\ %d
		10 &  & & T04\_035T01\_9000004622 (59430.59) & Pointed \\		
		11 & 90702326006	(59450.19) & & & \\
		12 & 90702326008	(59455.55) & & & \\
		13 &  & & T04\_046T01\_9000004680 (59457.06) & Pointed \\
		14 &  & & T04\_051T01\_9000004686 (59461.05) & Pointed \\		
		15 & 90702326010	(59465.67) & 4202230166	(59466.07) & &\\ %e
		16 & 90702326012 (59471.51) & 4202230171 (59471.43) &  &\\
	    &&&\\
		\hline
	\end{tabular}
\end{table*}

\section{OBSERVATIONS AND DATA REDUCTION}
\label{sec:redn}
We perform the present study based on the 2021 outburst of 4U $1543-47$ using \textit{NuSTAR, NICER} and \textit{AstroSat} observations over a period from 17 June 2021 (MJD 59382) to 14 September 2021 (MJD 59471). We considered all the \textit{NuSTAR} and \textit{AstroSat} observations in this period and used the \textit{NICER} observations which are simultaneous with \textit{NuSTAR}. The list of observations considered for this study is given in Table~\ref{tab:allobs}. There are a total of 16 epochs of observations consisting of ten \textit{NuSTAR} and six \textit{AstroSat} observations. Seven \textit{NuSTAR} epochs have simultaneous \textit{NICER} coverage also.

\subsection{\textit{NuSTAR} Data Reduction}
\label{sec:nustar} 
\textit{NuSTAR} \citep{2013ApJ...770..103H} has observed the source several times in the 2021 outburst. 
%For enormously bright sources, the detectors may experience pile-up issues while NuSTAR is devoid of such problems. 
\textit{NuSTAR} is devoid of pile-up issues and moreover, its good energy resolution in the energy coverage ($3-79$ keV) makes it suitable for the study of enormously bright sources like 4U $1543-47$. \textit{NuSTAR} consists two focal plane module telescopes (\textit{FPMA} and \textit{FPMB}), both are operating in $3-78$ keV band. The \textit{NuSTAR} data for the 2021 outburst is reduced using \texttt{HEASOFT v.6.29}, \texttt{NUSTARDAS pipeline v.2.1.1} and  \texttt{CALDB v.20211115}. For extremely bright sources, we set \texttt{statusexpr="STATUS==b0000xxx00xxxx000"}\footnote{\url{https://heasarc.gsfc.nasa.gov/docs/nustar/analysis/}} and set \texttt{saamode} to \texttt{strict} and \texttt{tentacle} to \texttt{yes}.
A circular region of radius 35 pixels centered on the brightest pixel is extracted as the source region and
as the background region, we also choose a 35 pixel circular region away from this. These files are used for generating science products such as the spectrum, background, lightcurve, Auxiliary Response File (ARF) and Response Matrix File (RMF), through the \texttt{NUPRODUCTS} task, independently for both \textit{FPMA} and \textit{FPMB}.
The spectra are grouped with a minimum of 50 counts per bin without any systematics.

\subsection{\textit{NICER} Data Reduction} 
\label{sec:nicer}
The X-ray Timing Instrument (\textit{XTI}) onboard \textit{NICER} \citep{2016SPIE.9905E..1HG} operates in $0.2-12$ keV band. \textit{NICER} has observed the source 4U $1543-47$ in almost every day during the 2021 outburst. We analyse \textit{NICER} data of the source between MJD 59382 and  MJD 59471 which is simultaneous with the \textit{NuSTAR} observations (Table~\ref{tab:allobs}). 
 The data is reduced using the tool \texttt{NICERDAS}\footnote{\url{https://heasarc.gsfc.nasa.gov/docs/nicer/nicer_analysis.html}} in HEASOFT v.6.29 with the 20210707 caldb version. 
There are 56 focal plane modules (FPMs) of \textit{NICER/XTI}.
We excluded FPM-14 and 34 in addition to the non-functional FPMs (FPM-11, 20, 22, and 60) due to increased noise levels. 
Since 4U $1543-47$ is extremely bright at the beginning of the outburst, the initial epochs (till $\sim$ MJD 59425) are affected by telemetry saturation. For such observations, a lower number of FPMs were kept active by the instrument team and we considered only the active detectors in the data reduction.
Level-2 standard calibration and filtering are done using \texttt{nicerl2} task and applied barycenter corrections through \texttt{barycorr} with \texttt{refframe="ICRS"}. Spectra is generated using \texttt{XSELECT (V2.4m)}. Lightcurve of the \textit{NICER} observation on MJD 59428.18  shows a flaring in the high energy band; therefore, the corresponding GTIs are excluded from the extraction.
The ARF and RMF files are generated for each observation based on the number of active detectors. The task \texttt{nibackgen3C50}\footnote{\url{https://heasarc.gsfc.nasa.gov/docs/nicer/tools/nicer_bkg_est_tools.html}}  \citep{2021arXiv210509901R} is used for creating background files. Finally, the source spectra are grouped with 25 photons per bin and a systematic uncertainty of 1 $\%$ is added to the spectra.

\begin{figure*}
	\centering
	\includegraphics[width=0.99\textwidth]{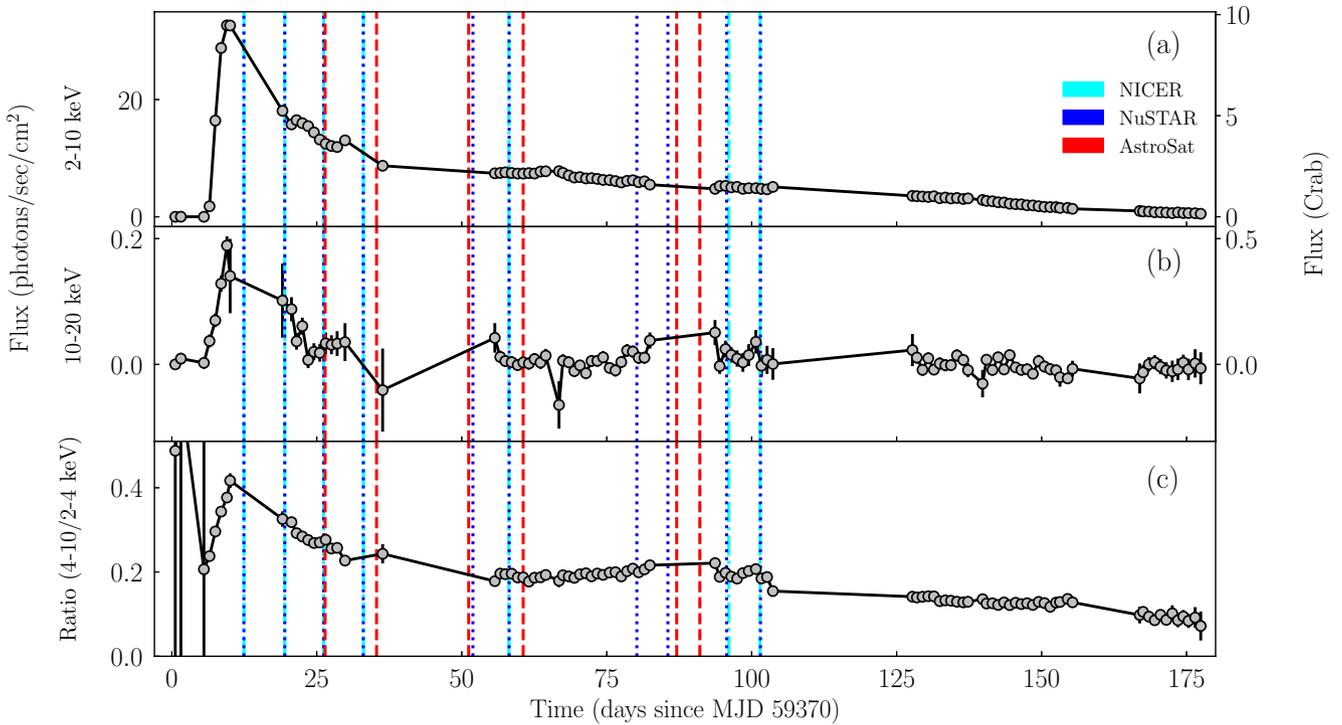} 
	\caption{\textit{MAXI/GSC} daily average lightcurve in the energy bands $\textbf{(a)}$ $2-10$ keV  and $\textbf{(b)}$ $10-20$ keV  with flux in  units of photons/sec/cm$^2$ and Crab in the left and right Y-axes respectively. The hardness ratio $\textbf{(c)}$ is defined by the flux in $4-10$ keV to $2-4$ keV. The \textit{NICER, NuSTAR} and \textit{AstroSat} observations for this study are marked using lines with the colours cyan, blue and red respectively.}
	\label{fig:lc-hr}
\end{figure*}
\subsection{\textit{AstroSat} Data Reduction}
\label{sec:astrosat}
The \textit{Soft X-ray Telescope} (\textit{SXT}) and \textit{Large Area X-ray Proportional Counter} (\textit{LAXPC}) on-board \textit{AstroSat} \citep{2016ApJ...833...27Y,YadavJAA2017} together observes the astronomical sources in wideband energy range (0.3$-$80 keV). \textit{AstroSat} has observed the 2021 outburst of 4U $1543$-$47$ during 6 different epochs. The first two of these observations are carried out with an offset of 40$^{\prime}$ since the source was too bright to have pointed observation \citep{2021ATel14749....1G}. We obtain Level-1 \textit{LAXPC} and Level-2 \textit{SXT} data of all six observations available at data archive hosted by \texttt{ISSDC}\footnote{\url{https://astrobrowse.issdc.gov.in/astro_archive/archive/Home.jsp}}. 

\textit{LAXPC} consists of three identical proportional counts namely \textit{LAXPC10}, \textit{LAXPC20} and \textit{LAXPC30}. However, for our analysis, we have used data from \textit{LAXPC20} alone because of its steady gain (see also \citealt{2021MNRAS.501.5457B,2021MNRAS.508.2447B,2022AdSpR..69..483B,2022MNRAS.514.6102P}). To extract the Level-2 \textit{LAXPC} data file i.e., source spectrum, lightcurve, RMF and background spectrum and lightcurve, we make use of latest version of single routine \textit{LAXPC} software \texttt{LaxpcSoftversion3.4.3}\footnote{\url{http://www.tifr.res.in/~astrosat\_laxpc/LaxpcSoft.html}} \citep{2022ApJS..260...40A}. Level-2 files are extracted from a single event and the top layer of \textit{LAXPC} unit to avoid the instrument effects at high energy. While the software generated \textit{LAXPC} response files are used for pointed observations, a 40$^{\prime}$ offset \textit{LAXPC} response file provided by the instrument is used for the offset observations (see also \citealt{2020MNRAS.497.1197B,2021MNRAS.501.6123K}).  

\textit{SXT} has observed the source in Photon Counting (PC) mode during all the epochs. The orbit-wise \textit{SXT} cleaned Level-2 event files are merged to get single event file for each observation using event merger python routine\footnote{\url{https://www.tifr.res.in/~astrosat_sxt/dataanalysis.html}} based on {\it Julia v 1.1.1}. The merged event file is then loaded into \texttt{XSELECT}, where we select single-pixel events by applying grade 0 filter to avoid optical data leakage \citep{2021JApA...42...77S,2022MNRAS.514.6102P}. From the \texttt{XSELECT} images, we find  that the first two offset observations have count rate $<40$ counts s$^{-1}$ and hence the corresponding spectra wouldn't have pileup issues. We select a circular region of radius 10$^{\prime}$ in the image to extract the source spectrum and lightcurve files.  
In all the pointed observations (see Table~\ref{tab:allobs}), we find the central region of the image to be very bright which could cause a pile-up effect. Therefore, source files are extracted from an annular region of the outer radius of 15$^{\prime}$ and inner radius of 2$^{\prime}$ for these observations. The standard \textit{SXT} background spectrum and instrument response file provided by the instrument team\footnote{\url{https://www.tifr.res.in/~astrosat_sxt/dataanalysis.html}} are used. ARF for the selected region is obtained from python-based tool \texttt{sxtarfmodule} provided by the \textit{SXT} team. Extracted \textit{SXT} and \textit{LAXPC} spectra are grouped to have 30 counts per bin in the first two observations and 20 counts per bin in the rest of the observations based on the source brightness. A systematics of 2\% \citep{2019MNRAS.487..928S,2022MNRAS.510.3019A} is applied for both \textit{SXT} and \textit{LAXPC} spectra.

\section{Outburst Profile and Hardness ratio}
\label{sec:lc-hr}
During the 2021 outburst of 4U $1543-47$, the flux reached the peak value within a few days of the commencement of the outburst. The outburst is monitored by multiple X-ray instruments. 
The \textit{MAXI/GSC}\footnote{\url{http://maxi.riken.jp/mxondem/}} daily lightcurve of the source is generated for two different energy bands, $2-10$ keV and $10-20$ keV and are plotted in Fig.~\ref{fig:lc-hr}. The MJD 59370 (05 June 2021) is defined as day 0 throughout the study and according to this, the outburst continues over $\sim$175 days.
The lightcurve reveals that the source is extremely luminous in low energies with a very high count rate (Fig.~\ref{fig:lc-hr}a), while the contribution to the luminosity in the high energy band (Fig.~\ref{fig:lc-hr}b) is an order of magnitude lower.
The highest value of flux in $2-10$ keV band is 32.67 photons/sec/cm$^2$ ($\sim$ 10 Crab) on day 9, whereas the same in $10-20$ keV is just 0.19 photons/sec/cm$^2$ ($\sim$ 0.5 Crab).
The source flux reached 11.65 Crab in $2-4$ keV, which is the highest value observed among the BH-XRBs. 
We define hardness ratio (HR) as the ratio of flux in $4-10$ keV to $2-4$ keV since beyond 10 keV the contribution is significantly low. The HR evolution (Fig.~\ref{fig:lc-hr}c) shows that the source mostly remains in the soft state during the outburst. The \textit{NICER, NuSTAR} and \textit{AstroSat} observations used in this study are marked by cyan, blue and red dashed lines, respectively. There is no simultaneous broadband observation in the rising phase of the outburst.

\begin{figure}
	\centering
		\includegraphics[width=0.33\textwidth,angle =-90]{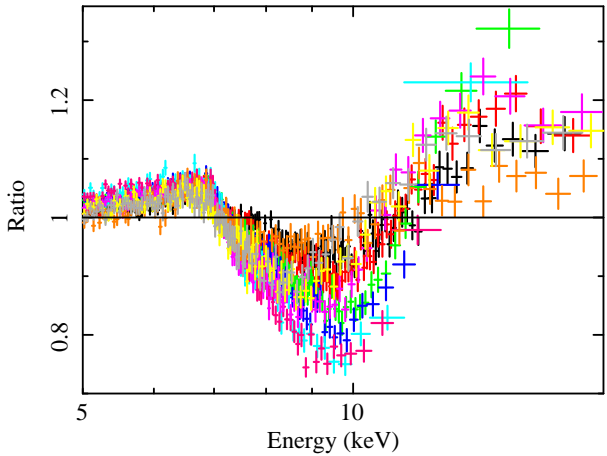} 
	\caption{The ratios (model to data) of the spectral fitting of \textit{NuSTAR} observations (Table \ref{tab:allobs}) using \textit{tbabs(diskbb+powerlaw)} model. The colours black, red, green, blue, cyan, pink, magenta, orange, yellow and grey represent the \textit{NuSTAR} epochs in ascending order (Table~\ref{tab:allobs}). A strong absorption feature exists between $\sim$ $8-11$ keV for all the epochs. The absorption depth increases up to Epoch 9 (pink in colour) and then decreases as the outburst progresses. The figure is zoomed around the absorption feature for better clarity.}
	\label{fig:nu-abso}
\end{figure}
\begin{figure*}
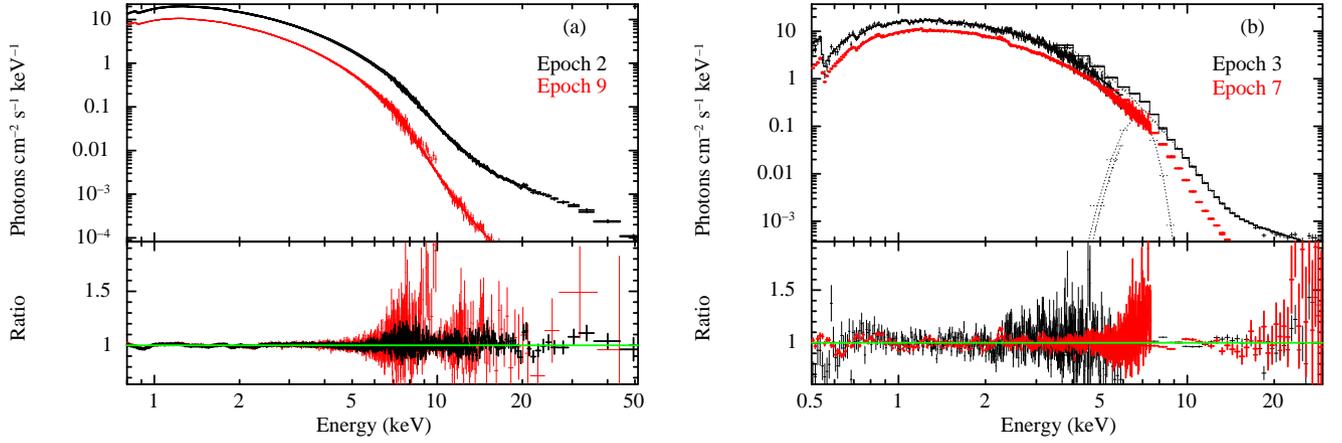
   
	\centering
		\includegraphics[width=0.332\textwidth,angle =-90]{f3a}
		\includegraphics[width=0.332\textwidth,angle =-90]{f3b}
	\caption{\textbf{(a)} Simultaneous \textit{NICER-NuSTAR} pair on Epoch 2 (black in colour) and Epoch 9 (red in colour) fitted with the model $tbabs(thcomp \times diskbb)edge \times gabs$. Epoch 2 spectrum is harder compared to that of Epoch 9.
	 \textbf{(b)} The \textit{AstroSat} spectra on Epoch 3 (black) and 7 (red) respectively modelled using $tbabs(thcomp \times diskbb) \times gabs$. We include an additional \textit{gauss} component for \textit{AstroSat} (Epoch 3) data. Both instruments show spectral changes during the outburst.}
	\label{fig:simut-pheno}
\end{figure*}

\section{Spectral Modelling and Results}
\label{sec:spec-result}
We studied the spectral properties of the 2021 outburst of 4U $1543-47$ from MJD 59382 (17 June 2021) to MJD 59471 (14 September 2021). All the three instruments,  \textit{NICER, NuSTAR} and \textit{AstroSat} have good coverage over this period. Table~\ref{tab:allobs} summarises the list of observations used in this work. In total, there are 16 epochs, comprising ten \textit{NuSTAR} and six \textit{AstroSat} (\textit{SXT-LAXPC}) observations. In addition, there are seven \textit{NICER} observations which are simultaneous with that of \textit{NuSTAR} and we used these pairs for \textit{NICER-NuSTAR} wideband spectral analysis.
We carried out phenomenological and reflection modelling of each \textit{NuSTAR} observations and extended that to wideband \textit{NICER-NuSTAR} and \textit{AstroSat} observations.

\subsection{Phenomenological Spectral Modelling}
\label{sec:pheno-modelling}
% nustar only
% ni-nu pairs
% astrosat
We used \texttt{HEASOFT v.6.29} and \texttt{XSPEC V12.12.0} package for the spectral modelling of \textit{NICER, NuSTAR} and \textit{AstroSat} data.  %\textcolor{teal}{ while \texttt{HEASOFT v.6.30} and %\texttt{XSPEC V12.12.1} for AstroSat}.
We have done spectral modelling of \textit{NICER-NuSTAR} and \textit{AstroSat} data to see the nature of the broad-band spectrum. The \textit{NICER} data below 0.8 keV shows large residuals; therefore, we used $0.8-10$ keV  for \textit{NICER} spectra and $4-60$ keV for \textit{NuSTAR} spectra. Spectra from both \textit{FPMA} and \textit{FPMB} telescopes of \textit{NuSTAR} give similar results. We present only \textit{FPMA} spectra throughout the study. We used $0.5-7$ keV for \textit{SXT} and $3-30$ keV for \textit{LAXPC} as significant data is not available beyond this range.

To accommodate the interstellar absorption, we used the \textit{tbabs} model which uses an equivalent hydrogen column density $n_H$ through the solar abundance table provided by \citet{2000ApJ...542..914W}. We initially modelled the \textit{NuSTAR} observations with \textit{tbabs(diskbb+powerlaw)} model. Here, \textit{diskbb} \citep{1984PASJ...36..741M} represents the multicolor blackbody spectrum from the accretion disc and \textit{powerlaw} employs the inverse Comptonization of the soft blackbody photons. We detected a broad absorption feature at $\sim$ $8-11$ keV in all epochs. The ratio of data to model of all the \textit{NuSTAR} observations are shown in Fig.~\ref{fig:nu-abso}. The different \textit{NuSTAR} epochs are shown in black, red, green, blue, cyan, pink, magenta, orange, yellow and grey colours in ascending order. It shows that the depth of absorption feature starts with a low value (black in colour) and then keeps on increasing as the outburst progress, reaching the maximum on Epoch 9 (pink in colour). Finally, the absorption depth decreases towards the end (grey in colour) of our study. 
We found a similar absorption feature in the \textit{AstroSat/LAXPC} data as well.

\definecolor{Gray}{gray}{0.9}
\definecolor{silver}{rgb}{0.75, 0.75, 0.75}
\begin{table*}
	\centering
	\caption{Wideband \textit{NICER-NuSTAR} simultaneous pairs and \textit{AstroSat} observations (highlighted with grey colour) using the model $tbabs(thcomp \times diskbb)edge \times gabs$ and $tbabs(thcomp \times diskbb) \times gabs$ respectively. The error values represent $90\%$ confidence interval. The \textit{NuSTAR} data on Epoch 8, 11 \& 12 are not included here as no simultaneous \textit{NICER} observations available. The bolometric ($0.5-100$ keV) observed flux and estimated luminosity for each epoch are also shown.} 
	\label{tab:pheno-ni-nu-param}
	\setlength{\tabcolsep}{2.2pt} % sets horizontal (column) spacing. Default value: 6pt
	\renewcommand{\arraystretch}{1.9} % sets vertical (row) spacing. Default value: 1
	\begin{tabular}{|c|c|c|c|c|c|c|c|c|c|c|c|c|c|c|}  
		\hline
	    \multirow{3}{*}{Epoch}& \multirow{3}{*}{$n_H$} & \multicolumn{9}{c}{Model} &  & \multirow{3}{*}{$\chi^2_{red}$} & \multirow{3}{*}{$F_{bol}$} & \multirow{3}{*}{$L_{bol}$}\\
	    \cline{3-12} 
	     & & \multicolumn{2}{c|}{\textit{diskbb}} & \multicolumn{3}{c|}{\textit{thcomp}} & \multicolumn{2}{c|}{\textit{edge} or \textit{gauss}$^*$} & \multicolumn{3}{c|}{\textit{gabs}} & & & \\
	     \cline{3-12}
	     & ($\times 10^{22}$ & $T_{in}$ & \textit{norm} &  $\Gamma$ & \textit{kTe} & \textit{cov$\_$frac} & \textit{line E} & \textit{D} or \textit{$\sigma$}  & \textit{line E} & $\sigma$ & \textit{strength} & & ($\times 10^{-8}$ & \\
	     & $cm^{-2}$) & (\textit{keV}) & ($ \times 10^3$) &  & (\textit{keV}) & ($ \times 10^{-2}$)  & (\textit{keV}) & (\textit{keV}) & (\textit{keV}) & (\textit{keV}) & (\textit{keV}) & & erg cm$^{-2}$ s$^{-1}$) & ($L_{Edd}$) \\
	    \hline
	
1& $0.470_{-0.002}^{+0.002}$ & $1.272_{-0.002}^{+0.002}$ & $7.96_{-0.05}^{+0.05}$ & $2.44_{-0.05}^{+0.06}$ & $20^f$ & $2.3_{-0.2}^{+0.2}$ & $7.16_{-0.05}^{+0.05}$ & $0.04_{-0.01}^{+0.01}$ & $10.0_{-0.1}^{+0.1}$ & $1.60_{-0.09}^{+0.09}$ & $0.84_{-0.08}^{+0.08}$ & 0.90 & $27.64_{-0.04}^{+0.04}$ & $1.52_{-0.14}^{+0.14}$ \\

2 & $0.458_{-0.002}^{+0.002}$ & $1.159_{-0.002}^{+0.002}$ & $6.07_{-0.04}^{+0.04}$ & $2.0_{-0.1}^{+0.1}$ & $11.6_{-1.9}^{+3.9}$ & $0.9_{-0.1}^{+0.2}$ & $7.19_{-0.07}^{+0.07}$ & $0.03_{-0.01}^{+0.01}$ & $9.9_{-0.1}^{+0.1}$ & $1.65_{-0.08}^{+0.07}$ & $1.2_{-0.1}^{+0.1}$ & 1.00 & $17.22_{-0.03}^{+0.02}$ & $0.95_{-0.09}^{+0.09}$ \\

\rowcolor{silver}
3 & $0.52_{-0.02}^{+0.02}$  & $0.99_{-0.01}^{+0.01 }$ & $15.4^{+1.5}_{-1.4}$ & $1.84^{+0.14}_{-0.12}$ & $20^{f}$ & $0.39^{+0.07}_{-0.06}$ & $6.77_{-0.15 }^{+0.15 }$ & $0.6^{f}$ & $7.35^{+0.34}_{-0.39}$ & $2^{f}$ & $1.73_{-0.47 }^{+0.47 }$ & $1.17$ & $19.5^{+0.1}_{-0.1}$ & $1.07^{+0.001}_{-0.001}$\\

4 & $0.461_{-0.002}^{+0.002}$ & $1.094_{-0.002}^{+0.002}$ & $5.82_{-0.05}^{+0.05}$ & $2.2_{-0.1}^{+0.1}$ & $20^f$ & $0.3_{-0.1}^{+0.1}$ & $7.27_{-0.07}^{+0.07}$ & $0.05_{-0.01}^{+0.01}$ & $10.4_{-0.1}^{+0.1}$ & $1.94_{-0.07}^{+0.08}$ & $2.2_{-0.1}^{+0.2}$ & 0.95 & $12.38_{-0.03}^{+0.02}$ & $0.68_{-0.06}^{+0.06}$ \\

5 & $0.458_{-0.002}^{+0.002}$ & $1.050_{-0.002}^{+0.002}$ & $5.77_{-0.04}^{+0.04}$ & $2.3_{-0.2}^{+0.2}$ & $20^f$ & $0.2_{-0.1}^{+0.1}$ & $7.28_{-0.05}^{+0.05}$ & $0.06_{-0.01}^{+0.01}$ & $10.7_{-0.1}^{+0.1}$ & $1.97_{-0.07}^{+0.08}$ & $2.7_{-0.2}^{+0.2}$ & 1.02 & $10.10_{-0.02}^{+0.01}$ & $0.56_{-0.05}^{+0.05}$ \\

\rowcolor{silver}
6 & $0.50_{-0.02 }^{+0.02 }$ & $0.97_{-0.02 }^{+0.02 }$ & $12.1^{+1.5}_{-1.4}$ & $2.3^{f}$ & $20^{f}$ & $0.3_{-0.02 }^{+0.02 }$ & $6.77_{-0.18 }^{+0.18 }$ & $0.6^{f}$ & $9.05^{+0.51}_{-0.55}$ & $2.5^{f}$ & $2.27^{+0.66}_{-0.68}$ & $1.27$ & $13.9^{+0.06}_{-0.05}$& $0.76^{+0.001}_{-0.001}$ \\

\rowcolor{silver}
7 & $0.54_{-0.01}^{+0.01 }$ & $0.98_{-0.01 }^{+0.01 }$ & $7.64_{-0.33}^{+0.33}$ & $2.26^{+0.06}_{-0.05}$ & $20^{f}$ & $0.2^{f}$ & - & - &  $10.7_{-0.2}^{+0.2}$ & $1.46^{+0.24}_{-0.18}$ & $1.56^{+0.25}_{-0.33}$ & 1.29 & $9.36^{+0.21}_{-0.20}$ & $0.51^{+0.004}_{-0.004}$ \\

9 & $0.451_{-0.003}^{+0.003}$ & $0.966_{-0.002}^{+0.002}$ & $5.60_{-0.05}^{+0.05}$ & $2.6_{-0.4}^{+0.5}$ & $20^f$ & $0.1_{-0.1}^{+0.2}$ & $7.44_{-0.08}^{+0.08}$ & $0.08_{-0.02}^{+0.02}$ & $10.5_{-0.1}^{+0.2}$ & $2.02_{-0.09}^{+0.09}$ & $3.3_{-0.2}^{+0.3}$ & 1.05 & $6.88_{-0.02}^{+0.02}$ & $0.38_{-0.04}^{+0.04}$ \\

\rowcolor{silver}
10 & $0.53_{-0.01 }^{+0.01 }$ & $0.96_{-0.01 }^{+0.01 }$ & $5.63^{+0.35}_{-0.31}$ & $1.92^{+0.30}_{-0.25}$ & $20^{f}$ & $0.11^{+0.05}_{-0.03}$ & - & - & $10.1_{-0.2 }^{+0.2 }$ & $1.56^{+0.20}_{-0.18}$ & $2.13^{+0.39}_{-0.32}$ & 1.28 & $7.59^{+0.04}_{-0.03}$ & $0.40^{+0.001}_{-0.001}$ \\

\rowcolor{silver}
13 & $0.52_{-0.01 }^{+0.01 }$ & $0.91_{-0.01 }^{+0.01 }$ & $7.48^{+0.24}_{-0.23}$ & $1.77_{-0.03 }^{+0.03 }$ & $20^{f}$ & $0.12_{-0.05 }^{+ 0.05}$ & - & - & $9.42_{-0.27 }^{+0.27 }$ & $0.75^{+0.30}_{-0.28}$ &  $0.47^{+0.13}_{-0.11}$& 1.10 & $6.59^{+0.06}_{-0.07}$ & $0.36^{+0.002}_{-0.003}$ \\

\rowcolor{silver}
14 & $0.48_{- 0.01}^{+0.01 }$ & $0.91_{-0.01 }^{+ 0.01}$ & $7.92^{+0.58}_{-0.53}$ & $1.94^{+0.02}_{-0.02}$ & $20^{f}$ & $0.4^{f}$ & - & - & $10.7^{+0.24}_{-0.21}$ & $1.34_{-0.17}^{+0.18}$ & $1.47_{-0.18}^{+0.18}$ & 1.31  & $6.22^{+1.89}_{-6.56}$  & $0.34^{+0.08}_{-0.29}$ \\

15 & $0.465_{-0.010}^{+0.004}$ & $0.904_{-0.004}^{+0.030}$ & $6.10_{-0.69}^{+0.09}$ & $2.09_{-0.04}^{+0.04}$ & $20^f$ & $1.6_{-0.1}^{+0.1}$ & $7.31_{-0.14}^{+0.08}$ & $0.10_{-0.03}^{+0.05}$ & $9.6_{-0.3}^{+0.3}$ & $2.1_{-0.4}^{+1.2}$ & $1.1_{-0.3}^{+1.1}$ & 1.27 & $5.66_{-0.02}^{+0.02}$ & $0.31_{-0.03}^{+0.03}$ \\

16 & $0.471_{-0.002}^{+0.002}$ & $0.897_{-0.001}^{+0.002}$ & $6.22_{-0.05}^{+0.05}$ & $2.04_{-0.06}^{+0.04}$ & $20^f$ & $1.6_{-0.1}^{+0.1}$ & $7.35_{-0.03}^{+0.06}$ & $0.12_{-0.03}^{+0.02}$ & $9.4_{-0.3}^{+0.4}$ & $1.5_{-0.3}^{+0.4}$ & $0.5_{-0.1}^{+0.1}$ & 1.00 & $5.39_{-0.01}^{+0.01}$ & $0.30_{-0.03}^{+0.03}$ \\
		\hline
	\end{tabular}			
	\begin{tablenotes}
			%\footnotesize   %% If you want them smaller like foot notes
			\item[b] $^*$ \textit{edge} is used for \textit{NICER-NuSTAR} pairs whereas \textit{gauss} component is used only for Epoch 3 \& 6 of \textit{AstroSat} data.\\
			%\item $^a$ Parameter uncertainty can't be estimated.\\
			\item $^f$ Frozen parameters

		\end{tablenotes}  
\end{table*}

 We used a partial covering fraction absorption model  \textit{pcfabs}\footnote{\url{https://heasarc.gsfc.nasa.gov/xanadu/xspec/manual/XSmodelPcfabs.html}} in XSPEC to check if this strong absorption feature at $\sim$ $8-11$ keV can be due to an intervening absorber, but it did not improve the fitting and the low energy residuals were high. We also tried several other models, for example, the thermal Comptonization model \textit{nthcomp} \citep{1996MNRAS.283..193Z} or \textit{thcomp} \citep{2020MNRAS.492.5234Z} in-place of \textit{powerlaw},  and \textit{diskbb} was replaced by \textit{kerrbb} \citep{2005ApJS..157..335L}. Various combinations of these models, fit to the data also showed the presence of the absorption feature at $\sim$ $8-11$ keV.
 Model combination with \textit{kerrbb} as the seed photon source did not provide a good fit; moreover, it failed to constrain the BH mass and spin. So, we prefer to use \textit{diskbb} model in combination with \textit{thcomp} which is an improved version of \textit{nthcomp}. The \textit{thcomp} is a convolution model which allows a variable fraction (parameter $cov\_frac$) of seed photon to Comptonize both up-scattering and down-scattering. Other parameters are the photon index ($\Gamma$) and electron temperature ($kT_e$) of the corona. 

 The observed absorption feature has a symmetrical profile and inclusion of a Gaussian absorption model \textit{gabs}\footnote{\label{note1}\url{ https://heasarc.gsfc.nasa.gov/xanadu/xspec/manual/node246.html}} fits the absorption feature well. The parameters of \textit{gabs} component are line energy (\textit{line E}, E$_g$), line width (\textit{$\sigma$}) and line depth (\textit{strength}). However, if we use a smeared absorption edge model, \textit{smedge} \citep{1994PASJ...46..375E} in XSPEC to compensate for the absorption feature, it results in an abnormally high value of absorption width due to the asymmetric nature of the model. 
 In addition, the \textit{NuSTAR} data show a weak presence of a Fe K$_\alpha$ absorption edge. Therefore, we also used an \textit{edge} component to improve the fit residual for \textit{NuSTAR}. The model parameters for the \textit{edge} component are the threshold energy of the absorption edge (\textit{edge E}, E$_e$) and the corresponding absorption depth (\textit{D}). Therefore, the final model for \textit{NICER-NuSTAR} data is  $tbabs(thcomp \times diskbb)edge \times gabs$ (model M1, hereafter).
 In contrast, the absorption edge feature was not visible in the \textit{AstroSat/LAXPC} data, possibly due to the low spectral resolution of \textit{LAXPC}. Therefore, the model M1 for \textit{AstroSat} data becomes $tbabs(thcomp \times diskbb)\times gabs$. However, the \textit{AstroSat} data of Epoch 3 \& 6 show the presence of a weak Fe K$_\alpha$ emission line feature in the residual, and we include a \textit{gauss} model component for these two epochs of \textit{AstroSat} data. 
%  Where the \textit{gabs} and \textit{edge} components are applied on the reprocessed spectrum. 

All the seven \textit{NICER-NuSTAR} simultaneous pairs and six \textit{AstroSat} observations are fitted with the model M1. The Fig.~\ref{fig:simut-pheno}a and Fig.~\ref{fig:simut-pheno}b show the wideband spectra of \textit{NICER-NuSTAR} (Epoch 2 \& 9) and \textit{AstroSat} (Epoch 3 \& 7) respectively modelled using M1. In each case, the spectrum in the red colour is relatively softer than that of the black colour; therefore, both figures illustrate moderate spectral changes during the outburst decay. 

\begin{figure}
	\centering 
		\includegraphics[width=0.47\textwidth]{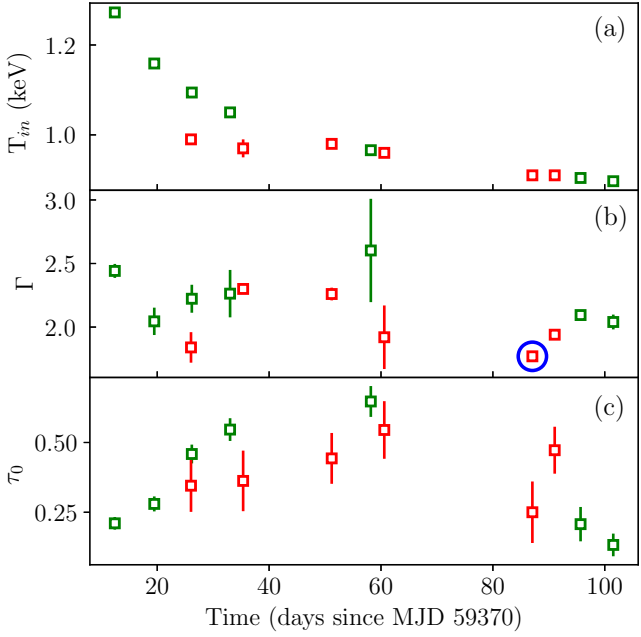} 
	\caption{Evolution of \textbf{(a)} T$_{in}$ and \textbf{(b)} $\Gamma$ from simultaneous \textit{NICER-NuSTAR} pairs (green in colour) and \textit{AstroSat} observations (red in colour) fitted using the model M1. We calculate the (\textbf{c}) optical depth ($\tau_0$) of the absorber from \textit{gabs} components for each epoch. The $\Gamma$ of Epoch 13 marked with a blue circle carries a special signature that is discussed in \S\ref{sec:discuss}.}
	\label{fig:plot-pheno}
\end{figure}

The goodness of the fit is determined using $\chi^2$ statistics. The reduced $\chi^2$ ($\chi^2_{red}$) varies between $0.9-1.3$. All the parameters estimated from the wideband phenomenological modelling are presented in Table~\ref{tab:pheno-ni-nu-param}. The parameter uncertainties are calculated within the $90\%$ confidence range. Note that the \textit{NuSTAR} observations on Epoch
8, 11 \& 12 (Table~\ref{tab:allobs}) are not included here as no simultaneous \textit{NICER} observations available. The $n_H$ is left free and it varies between ($0.45-0.54$) $\times$ 10$^{22}$ cm$^{-2}$. We aim to find out the evolution of various parameters with the progress of the outburst. Fig.~\ref{fig:plot-pheno} gives the variation of the inner disk temperature $T_{in}$, photon index $\Gamma$ and optical depth ($\tau_0$) with time. Here, points in green and red colour represent the parameter value estimated from \textit{NICER-NuSTAR} and \textit{AstroSat} spectral modelling respectively.

We find that the inner disk temperature (Fig.~\ref{fig:plot-pheno}a and Table~\ref{tab:pheno-ni-nu-param}) monotonically decreases throughout the outburst decay.
%due to the reduction of mass accretion rate $\dot M$. 
%\textcolor{red}{The \textit{diskbb norm} (Table~\ref{tab:pheno-ni-nu-param}) shows a decreasing trend; therefore, the inner disk radius $r_{in}$ marginally moves inward since \textit{diskbb norm} $\propto$ $r_{in}^2$.}
%\textcolor{red}{However, the decrease in $r_{in}$ could not prevent the drop in $T_{in}$ due to the gradual decline in $\dot M$ as $T_{in}$ $\propto$ $\dot M^{1/4}$ $r_{in}^{-3/4}$ \citep{frank_king_raine_2002}.}
The evolution of the \textit{diskbb norm} (Table~\ref{tab:pheno-ni-nu-param}) estimated using \textit{NICER-NuSTAR} decreases till Epoch 9 and a reverse trend is observed for later epochs. The \textit{diskbb norm} from the \textit{AstroSat} data also shows a similar pattern though \textit{AstroSat} values are higher than that of \textit{NICER-NuSTAR}.
The variation of photon index $\Gamma$ is presented in Fig.~\ref{fig:plot-pheno}b. The value of $\Gamma$ estimated from \textit{AstroSat} data (red square) differs from that of \textit{NICER-NuSTAR} pairs (green square). This can be due to the non-availability of the high energy contribution (beyond 30 keV) in the \textit{LAXPC} data. From \textit{NICER-NuSTAR} fitting, $\Gamma$ varies between $2-2.6$, and it shows spectral softening till Epoch 9. 
We wanted to estimate the electron temperature ($kT_e$) of the corona using the \textit{thcomp} model. 
But the broadband spectral fitting could not constrain the value of $kT_e$, except for Epoch 2, for which we obtained $kT_e=$ 11.6$_{-1.9}^{+3.9}$ keV (Table~\ref{tab:pheno-ni-nu-param}). For all the remaining epochs, we freeze $kT_e$ at 20 keV. 
Only a tiny fraction, \textit{cov$\_$frac} < 3 \% (Table~\ref{tab:pheno-ni-nu-param}), of the soft photons Comptonized in the corona. It gradually decreases till Epoch 10 and increases afterwards. This behaviour is consistent with the spectral softening trend shown in Fig.~\ref{fig:plot-pheno}b.

The broad absorption feature at $\sim$ 8$-$11 keV in the spectrum is well represented by the \textit{gabs} model. The \textit{strength} shows an increasing trend and reaches the maximum in Epoch 9 and declines beyond that. We calculate the optical depth ($\tau_0$) associated with the \textit{gabs} component using \textit{gabs strength} and $\sigma$ as $\tau_0=$ \textit{strength}$/  \sigma \sqrt{2 \pi}$. The evolution of $\tau_0$ is shown in Fig.~\ref{fig:plot-pheno}c, which shows that the absorption optical depth increases and reaches a maximum on Epoch 9 and then decreases. The dynamic behaviour of the absorption \textit{strength} seems interesting. We attempt to characterize the strong and dynamic absorption features in \S\ref{sec:absorption}.
The evolution of the \textit{edge} component is listed in Table~\ref{tab:pheno-ni-nu-param}. We discuss a possible connection between \textit{edge} and \textit{gabs} components in \S\ref{sec:discuss}. 

We estimate the observed bolometric flux ($F_{bol}$) in $0.5-100$ keV with uncertainty in $90\%$ confidence interval from the wideband simultaneous spectral data, which is also shown in Table~\ref{tab:pheno-ni-nu-param}.
Corresponding bolometric luminosity ($L_{bol}$) of the source is also calculated by assuming the distance to the source as $7.5 \pm 0.5$ kpc \citep{2004MNRAS.354..355J}. The Eddington luminosity of the source is $L_{Edd}=1.22 \pm 1 \times 10^{39}$ ergs s$^{-1}$ with an assumed BH mass of $9.4 \pm 1$ M$_\odot$ \citep{2006MNRAS.371.1334R}. It can be seen from Table~\ref{tab:pheno-ni-nu-param} that the luminosity of the source exceeds the $L_{Edd}$ at the peak (Epoch 1 is close to the peak) of the outburst, and the luminosity decreases gradually with the decay of the outburst. 

\begin{figure}
	\centering
		\includegraphics[width=0.48\textwidth]{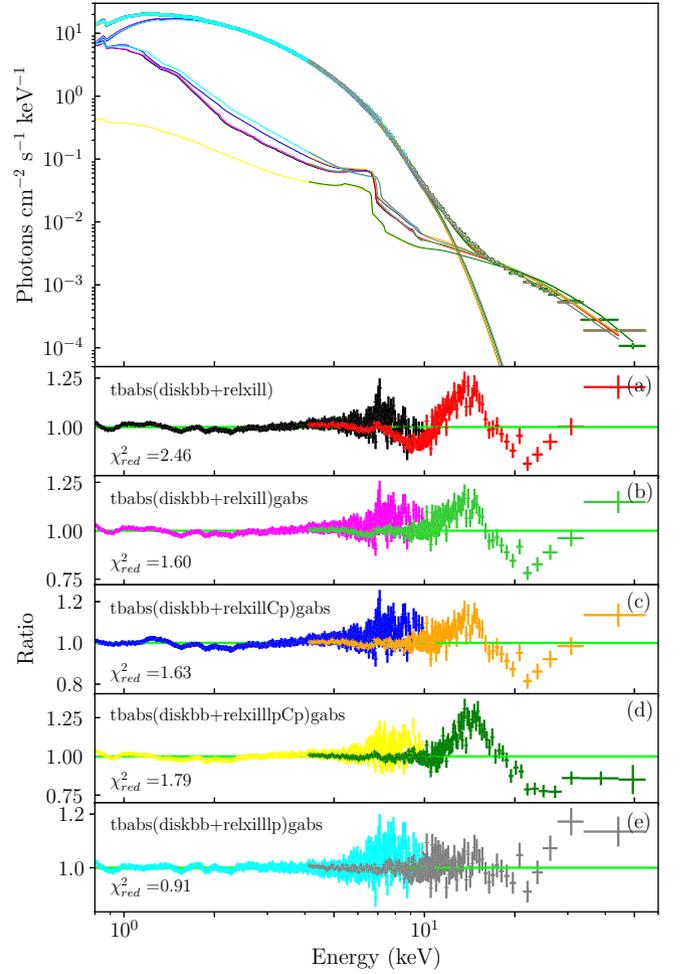} 
	\caption{Unfolded spectrum (top panel) of simultaneous \textit{NICER-NuSTAR} pair on Epoch 2 and ratio of the model to the data using various reflection models \textbf{(a)} \textit{tbabs(diskbb+relxill)}, \textbf{(b)} \textit{tbabs(diskbb+relxill)gabs}, \textbf{(c)} \textit{tbabs(diskbb+relxillCp)gabs}, \textbf{(d)} \textit{tbabs(diskbb+relxilllpCp)gabs} and \textbf{(e)} \textit{tbabs(diskbb+relxilllp)gabs}. The model and the value of  $\chi^2_{red}$ are mentioned at the top left and bottom left corners, respectively, in each case.}
	\label{fig:refl-trial}
\end{figure}

\subsection{Spectral Modelling for Reflection Studies}
\label{sec:reflec-modelling}

%\citep{2014ApJ...782...76G}
To understand the reflection features in the spectra of 4U $1543-47$, we use the relativistic reflection model RELXILL\footnote{\url{http://www.sternwarte.uni-erlangen.de/~dauser/research/relxill/index.html}}v1.4.3. 
Different flavours of the RELXILL model are tried. The unfolded \textit{NICER-NuSTAR} spectra and data-to-model ratios using various reflection models are shown in Fig.~\ref{fig:refl-trial} for Epoch 2.
We started with the model \textit{tbabs(diskbb+relxill)} which gives a $\chi^2_{red}$ of $2.46$. The data-to-model ratio (Fig.~\ref{fig:refl-trial}a) shows that the absorption feature at $\sim 8-11$ keV in the spectrum cannot be fitted by the reflection model. We added the absorption model \textit{gabs} with this, and the model \textit{tbabs(diskbb+relxill)gabs} (Fig.~\ref{fig:refl-trial}b) improves the residual but still the $\chi^2_{red}$ for this combination is $1.6$.
 Then, we replaced the model \textit{relxill} with \textit{relxillCp} (Fig.~\ref{fig:refl-trial}c) where a thermal Comptonizing continuum is assumed for the illuminating flux. This combination has $\chi^2_{red}$ $=1.63$, which is also unacceptable. Then, we used \textit{relxilllpCp} (Fig.~\ref{fig:refl-trial}d) as the reflection model, where a lamp-post (\textit{lp}) geometry is assumed for the corona. 
 In all the ‘\textit{lp}’ flavours of the RELXILL model, the inner disk is illuminated by a point-like corona situated at a height ‘$h$’ from the disk surface on the axis of rotation.
The modelling with \textit{relxilllpCp} also resulted in a large residual, with $\chi^2_{red}=1.79$. We replaced \textit{relxilllpCp} with \textit{relxilllp} (Fig.~\ref{fig:refl-trial}e) where the illuminating flux is modelled as a \textit{powerlaw} with a high-energy cutoff just like the \textit{relxill}.
 For this trial, we got a reasonable fit with $\chi^2_{red}=0.91$, and we decided to proceed with the model combination, \textit{tbabs(diskbb+relxilllp)gabs} (model M2, hereafter) as the final model to study the reflection features in the spectra. 
 Note that, no additional \textit{edge} or \textit{gauss} component is required in the reflection modelling. %\textcolor{red}{However, the reflection model could not manage the broad absorption feature $\sim 8-11$ keV and an additional \textit{gabs} component is required to fit the data.--repeatative} 

Using the model M2, we did the spectral fitting of the simultaneous \textit{NICER-NuSTAR} pairs and \textit{AstroSat} observations. Here, $n_H$ is a free parameter.  
Inner and outer disk radii are frozen at the innermost stable circular orbit $R_{ISCO}$ and 400 $r_g$ (where $r_g \equiv GM/c^2$, the gravitational radius of the BH), respectively.
The \textit{powerlaw} cutoff energy, $E_{cut}$, is fixed at 60 keV since it is hitting the upper limit.  
All other parameters of \textit{relxilllp} are kept free; the lamp-post height $h$ (in units of $r_g$), inclination angle of the system $\theta $ (in degree), $\Gamma$ of the incident radiation, ionization parameter $log$ $\xi$ (erg cm s$^{-1}$), iron abundance ($A_{Fe}$) of the accretion disk in terms of the solar abundance $A_{Fe,\odot}$, and the reflection fraction $R_f$. Here, $R_f$ is defined as the ratio of the primary photon flux illuminating the disk to that reach the observer at infinity \citep{2016A&A...590A..76D}. 
We wish to estimate the spin parameter, $a_\ast$, of the system, but it is found to hit the upper limit for almost all the epochs. 
%We could able to estimate it only for Epoch 2 by freezing the inclination $\theta$,  $log$ $\xi$ and $A_{Fe}$ to their best fitted values, and we obtained $a_\ast= 0.5_{-0.4}^{+0.3}$. This agrees with the result reported by \citet{2014ApJ...793L..33M} and \citet{2020MNRAS.493.4409D}. 
Based on previous studies \citep{2014ApJ...793L..33M, 2020MNRAS.493.4409D}, we freeze $a_\ast=0.4$ for this study.

 The estimated reflection model (M2) parameters are listed in Table~\ref{tab:refl-ni-nu-param}. The errors represent 90$\%$ confidence interval. The evolution of $T_{in}$ follows the same trend as that observed in the phenomenological spectral modelling (Table~\ref{tab:pheno-ni-nu-param}). The value of $\Gamma$ varies between $2-3.3$ and appears slightly steeper than that of phenomenological modelling (Table~\ref{tab:pheno-ni-nu-param}). It may be due to the additional low energy contribution from the reflection component over the \textit{diskbb} component. However, reflection modelling also shows spectral softening till Epoch 9. The \textit{NICER-NuSTAR} results suggest that 4U $1543-47$ is a low inclination system with $\theta$ varies between $\sim 32^\circ - 40^\circ$. We could not constrain the inclination angle from \textit{AstroSat} data and freeze it to 40$^\circ$.

\begin{figure}
	\centering 
		\includegraphics[width=0.47\textwidth]{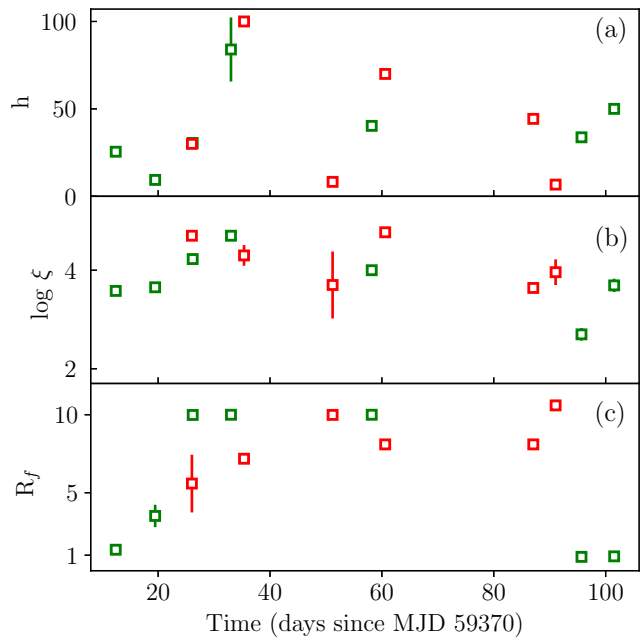} 
	\caption{Evolution of \textbf{(a)} $h$, \textbf{(b)} log $\xi$ and \textbf{(c)} R$_f$ from reflection modelling of simultaneous \textit{NICER-NuSTAR} pairs (green in colour) and \textit{AstroSat} observations (red in colour) using the model $tbabs(diskbb+relxilllp)gabs$. See text for details.}
	\label{fig:plot-refl}
\end{figure}

 The evolution of few important model parameters (\textit{h}, log $\xi$ and R$_f$) are shown in Fig.~\ref{fig:plot-refl} for better presentation. Simultaneous \textit{NICER-NuSTAR} pairs and \textit{AstroSat} observations are marked in green and red colour, respectively. We could not estimate the uncertainty of \textit{h} in most of the epochs. The evolution of \textit{h} (Fig.~\ref{fig:plot-refl}a) indicates that the primary source is moving away from the BH till Epoch 6 and then gradually coming closer to the central object.
The ionization structure of the disk is established through the parameter log $\xi$. Its value gradually increases (Fig.~\ref{fig:plot-refl}b) and reaches the maximum around Epoch 10 and then gradually decreases. The high value of log $\xi$ (>3) suggests a highly ionized disc material throughout the outburst. We could estimate iron abundance A$_{Fe}$ for Epoch 1, 2, 4 and 5, and it hits the upper limit during the rest of the epochs. Our study reveals an overabundance (3.6$-$10 $A_{Fe,\odot}$) of iron in the disk. The reflection fraction R$_f$ is estimated well at the first three and last two epochs only. Fig.~\ref{fig:plot-refl}c suggests that the fraction of primary photons reaching the disk increases till Epoch 9, and it decreases afterwards. The \textit{gabs strength} shows a similar behaviour found in the phenomenological modelling. We have discussed more on this result in \S\ref{sec:absorption}.

%We aim for the spin estimation of the system, which we have frozen at $0.4$ as we discussed before. Now we thaw it and checked, the values are either hitting the limits or not able to estimate the uncertainty for almost all the pairs. 

\begin{table*}
	\centering
	\caption{Reflection modelling of \textit{NICER-NuSTAR} simultaneous pairs and \textit{AstroSat} observations (highlighted with grey colour) using the model \textit{tbabs(diskbb+relxilllp)gabs}. The error values represent $90\%$ confidence interval. The \textit{NuSTAR} data on Epoch 8, 11 \& 12 are not included here as no simultaneous \textit{NICER} observations available.}
	\label{tab:refl-ni-nu-param}
	\setlength{\tabcolsep}{1.7pt} % sets horizontal (column) spacing. Default value: 6pt
	\renewcommand{\arraystretch}{1.9} % sets vertical (row) spacing. Default value: 1
	\begin{tabular}{|c|c|c|c|c|c|c|c|c|c|c|c|c|c|c|}  
		\hline
	\multirow{3}{*}{Epoch} & \multirow{3}{*}{$n_H$} & \multicolumn{12}{c|}{Model} & \multirow{3}{*}{$\chi^2_{red}$} \\
	\cline{3-14} 
	& &\multicolumn{2}{c|}{\textit{diskbb}}  & \multicolumn{7}{c|}{\textit{relxilllp}} &\multicolumn{3}{c|}{\textit{gabs}} &\\
	\cline{3-14} 
    &($\times 10^{22}$ &$T_{in}$ & \textit{norm} & \textit{h} & $\theta$  & $\Gamma$ & $log$ $\xi $ & $A_{Fe}$ & $R_f$ &  \textit{norm} & \textit{line E} & $\sigma$ & \textit{strength} & \\
    &$cm^{-2}$) &\textit{(keV)} & ($ \times 10^3$) & $(GM/c^2)$ & (\textit{deg}) & & (\textit{erg cm s$^{-1}$}) & ($A_{Fe,\odot}$) & &  ($\times 10^{-3}$)&(\textit{keV}) & (\textit{keV})& (\textit{keV}) & \\ 
		\hline
1 & $0.49_{-0.01}^{+0.01}$ & $1.276_{-0.001}^{+0.001}$ & $6.31_{-0.03}^{+0.03}$ & $25.50^a$ & $32.7_{-2.9}^{+3.4}$ & $2.81_{-0.03}^{+0.03}$ & $3.58_{-0.07}^{+0.09}$ & $ 8.5_{-1.3}^{+1.2}$ & $1.4_{-0.2}^{+0.1}$ & $282.7_{-46.1}^{+84.3}$ & $9.9_{-0.1}^{+0.1}$ & $2.05_{-0.03}^{+0.1}$ & $1.1_{-0.1}^{+0.1}$ & 0.80 \\

2 & $0.70_{-0.05}^{+0.04}$ & $1.151_{-0.005}^{+0.006}$ & $6.3_{-0.2}^{+0.2}$ & $9.4^a$ & $36.3_{-1.8}^{+1.6}$ & $2.94_{-0.08}^{+0.04}$ & $3.65_{-0.08}^{+0.05}$ & $7.7_{-1.1}^{+1.2}$ & $3.5_{-0.7}^{+0.9}$ & $200.4_{-53.9}^{+47.8}$ & $9.8_{-0.1}^{+0.1}$ & $1.94_{-0.09}^{+0.09}$ & $1.3_{-0.2}^{+0.2}$ & 0.91 \\
\rowcolor{silver} 
3 & $0.46^{+0.02}_{-0.01}$ & $1.01_{-0.006 }^{+0.006}$ & $11.7^{+0.76}_{-0.44}$ & $30^{f}$ & $40^{f}$ & $3.0^{f}$ & $4.7^{b}$ & $9.37^{b}_{-0.40}$ & $5.59^{+1.85}_{-3.48}$ & $0.32_{-0.08}^{+0.08}$ & $9.37^{+0.27}_{-0.13}$ & $0.75^{+0.37}_{-0.64}$ & $0.48^{+0.03}_{-0.19}$ &  1.37 \\
 
4 & $0.683_{-0.03}^{+0.004}$ & $1.081_{-0.001}^{+0.001}$ & $6.14_{-0.04}^{+0.008}$ & $30.5_{-2.0}^{+24.3}$ & $35.7_{-2.7}^{+2.2}$ & $3.23_{-0.06}^{+0.01}$ & $4.23_{-0.04}^{+0.05}$ & $5.8_{-0.7}^{+0.8}$ & $10.0^a$ & $41.1_{-2.0}^{+20.3}$ & $10.04_{-0.07}^{+0.04}$ & $1.92_{-0.06}^{+0.04}$ & $1.99_{-0.08}^{+0.2}$ & 0.83 \\

5 & $0.67_{-0.03}^{+0.01}$ & $1.044_{-0.001}^{+0.001}$ & $5.83_{-0.04}^{+0.01}$ & $84.0_{-18.3}^{+56.6}$ & $39.6_{-3.1}^{+2.6}$ & $3.385_{-0.05}^{+0.006}$ & $4.7^b$ & $7.1_{-0.7}^{+0.6}$ & $10.0^a$ & $33.6_{-2.5}^{+3.7}$ & $10.33_{-0.03}^{+0.03}$ & $2.06_{-0.05}^{+0.03}$ & $2.76_{-0.06}^{+0.02}$ & 0.83 \\

\rowcolor{silver}
6 & $0.49^{+0.01}_{-0.02}$ & $0.99^{+0.07}_{-0.01}$ & $9.23^{+1.46}_{-0.19}$ & $100^{f}$ & $40^{f}$ & $3^{f}$ & $4.30_{-0.21}^{+0.21}$ & $10^{b}$ & $7.18^{a}$ & $7.11^{+1.95}_{-4.1}$ & $10.0^{+0.13}_{-0.14}$ & $1.30_{-0.13}^{+0.13}$ & $1.39^{+0.28}_{-0.11}$  & 1.30 \\

\rowcolor{silver}
7 & $0.53^{+0.03}_{-0.01}$ & $0.99_{-0.003}^{+0.003}$ & $6.94^{+0.16}_{-0.27}$ & $8.26^{a}$ & $40^{f}$ & $2.46^{+0.36}_{-0.25}$ & $3.70^{+0.75}_{-0.68}$ & $10^{b}$ & $9.99^{a}$ & $5.75^{+0.55}_{-0.69}$ & $10.5^{+0.16}_{-0.14}$ & $1.45^{+0.20}_{-0.17}$ & $1.65^{+0.30}_{-0.25}$ & 1.44 \\

9 & $0.496_{-0.007}^{+0.01}$ & $0.958_{-0.001}^{+0.001}$ & $
 5.85_{-0.04}^{+0.04}$ & $40.29^a$ & $40^f$ & $3.17_{-0.06}^{+0.1}$ & $4.7^b$ & $10^b$ & $10^f$ & $4.0_{-1.0}^{+2.3}$ & $10.11_{-0.08}^{+0.08}$ & $1.86_{-0.06}^{+0.06}$ & $3.1_{-0.2}^{+0.2}$ &
 1.10\\
 
\rowcolor{silver}
10 & $0.44^{+0.03}_{-0.01}$ & $0.98_{-0.004}^{+0.004}$ & $4.95^{+0.10}_{-0.90}$ &  $70^{f}$& $40^{f}$ & $2.87^{+0.33}_{-0.24}$ & $4.7^{b}$ & $10^{b}$ & $8.10^{a}$ & $1.33^{+0.28}_{-0.12}$ & $10.2^{+0.13}_{-0.12}$ & $1.78^{+0.16}_{-0.19}$ & $2.76^{+0.36}_{-0.38}$  & 1.19 \\

\rowcolor{silver}
13 & $0.47_{-0.01}^{+0.01}$ & $0.94^{+0.002}_{-0.003}$ & $5.90^{+0.08}_{-0.09}$ & $44.3^{a}$ & $40^{f}$ & $2.25^{f}$ & $3.64^{+0.09}_{-0.18}$ & $3.56^{+0.56}_{-0.46}$ & $8.10^{a}$ & $1.11^{+0.07}_{-0.16}$ & $9.22^{+0.21}_{-0.18}$ & $0.84^{+0.20}_{-0.21}$ & $0.74^{+0.10}_{-0.13}$ & 1.11 \\

\rowcolor{silver}
14 & $0.48_{-0.01}^{+0.01}$ & $0.91_{-0.002}^{+0.002}$ & $7.95^{+0.21}_{-0.12}$ & $6.69^{a}$ & $40^{f}$ & $2.61^{+0.08}_{-0.19}$ & $3.96^{+0.07}_{-0.26}$ & $10^{b}$ & $10.6^{a}$ & $0.14_{-0.01}^{+0.01}$ & $10.1_{-0.13}^{+0.13}$ & $1.03^{+0.12}_{-0.13}$ & $1.25_{-0.14}^{+0.14}$ & 1.24 \\

15 & $0.475_{-0.003}^{+0.004}$ & $0.906_{-0.001}^{+0.001}$ & $5.91_{-0.04}^{+0.04}$ & $33.74^a$ & $38.0_{-4.4}^{+8.5}$ & $2.07_{-0.06}^{+0.06}$ & $2.7_{-0.3}^{+0.1}$ & $10.0^b$ & $0.9_{-0.2}^{+0.2}$ & $4.9_{-0.7}^{+1.5}$ & $9.1_{-0.1}^{+0.1}$ & $1.5_{-0.2}^{+0.1}$ & $1.0_{-0.1}^{+0.1}$ & 1.12 \\

16 & $0.482_{-0.004}^{+0.004}$ & $0.904_{-0.001}^{+0.001}$ & $5.63_{-0.05}^{+0.05}$ & $50.01^a$ & $36.7_{-5.5}^{+9.4}$ & $2.05_{-0.05}^{+0.05}$ & $3.7_{-0.2}^{+0.1}$ & $10.0^b$ & $0.9_{-0.1}^{+0.2}$ & $4.2_{-0.6}^{+0.8}$ & $8.7_{-0.1}^{+0.1}$ & $1.49_{-0.08}^{+0.08}$ & $1.2_{-0.2}^{+0.2}$ & 
0.87 \\
%& & & & & & & & & & & & & & \\
		\hline
	\end{tabular}
			\begin{tablenotes}
			%\footnotesize   %% If you want them smaller like foot notes
			\item[a] $^a$ Parameter uncertainty can't be estimated.\\
			\item[] $^b$ Parameter hits the boundary.\\
			\item[] $^{f}$ Frozen parameters.
		\end{tablenotes}  
\end{table*}

Very recently, the RELXILL model (version v2.2) has undergone some modifications by considering the effect of returning radiation in the calculation of reflected flux. Particularly in \textit{relxilllpCp}, where the effects of returning radiation, the density profile and ionization gradient of the disk and the velocity of the primary source are also included. However, the velocity of the primary source and the effects of returning radiation are the new parameters added to the \textit{relxilllp} model. We applied this modified \textit{relxilllp} model to the broadband \textit{NICER-NuSTAR} data but could not constrain the source velocity. 
Also, we tried to estimate the parameters using the v2.2 flavour of \textit{relxilllpCp}. We fitted all the wideband \textit{NICER-NuSTAR} observations using the model \textit{tbabs(diskbb+relxilllpCp)gabs}. But, we could not constrain most of the parameters since the number of free parameters is very large. Therefore, we need to essentially freeze all the new parameters introduced in the updated version, and RELXILL v2.2 does not bring any improvement in the result.
%Also, For example, the density (log N) of the accretion disk hits the upper boundary in most of the . The $log$ N shows an increasing trend which is not allowed since the mass accretion rate is decreasing throughout the outburst.
 
%According to this scenario, the absorption feature is strongest for Epoch 8 in individual \textit{NuSTAR} fitting and Epoch 5 in wideband fitting whereas in our analysis (using RELXILL v1.4.3) Epoch 9 is showing the strongest absorption feature in both cases.
%The reflection fraction also shows the similar trend. The value of $\theta$ varies between   $28^\circ-43^\circ$ for wideband fitting. Lamp-post height is very showing random behaviour. $log$ $\xi$ is not showing any trend for both cases.

\subsection{Absorption Features in the Spectra of 4U 1543$-$47}
\label{sec:absorption}

The wideband spectral analysis of 4U $1543-47$ reveals the presence of a very strong absorption feature (Fig.~\ref{fig:nu-abso}, Table~\ref{tab:pheno-ni-nu-param}, Table~\ref{tab:refl-ni-nu-param}) whose strength changes throughout the outburst. 
 We use \textit{gabs} model to characterize the absorption feature in phenomenological and reflection modelling. The \textit{gabs strength} estimated from both methods follow the same trend; getting more stronger as the outburst progresses and reaches the maximum value on Epoch 9, then declines gradually.

In general, the absorption features in the spectrum can be due to multiple reasons like the presence of  obscuring cloud in the line-of-sight, occultation due to the companion star, strong accretion disk-wind %\citep{2008ApJ...680.1359M}, 
and/or the stellar wind from the companion \citep{2008ApJ...680.1359M,2008MNRAS.386..593S,2020A&A...639A..13K}. We have discarded the chances of absorption due to obscuring cloud in the line-of-sight by fitting the data with the partial covering fraction model \textit{pcfabs} and found no improvement in the fitting. 
  If the absorption feature is produced by the occultation or stellar wind of the binary companion, the  features must show some orbital variations.
 Precise diagnostic of the orbital variations
 provide significant insight into the understanding of the nature and origin of the absorption features. Since 4U $1543-47$ is a low inclination system ($\theta \sim 32^\circ-40^\circ$), the expected orbital variation of the absorption feature, if any, will be weak. Therefore, we avoid using multi-instrument data to check the orbital variation of the absorption feature, as the differences in the estimated parameters between instruments may screw up the variation. We use only the \textit{NuSTAR} observations for this purpose.
 
\begin{figure}
	\centering
		\includegraphics[width=0.325\textwidth,angle =-90]{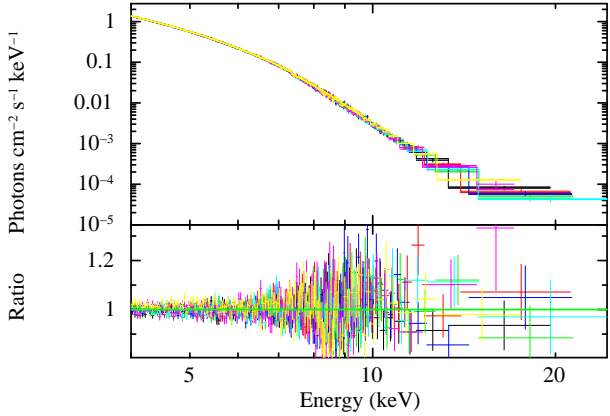} 
	\caption{Folded spectra of different patches in \textit{NuSTAR} observation of Epoch 9 using the model $tbabs(thcomp \times diskbb)edge \times gabs$. The parameters, except the \textit{gabs strength}, are tied between the patches. See text for details.}
	\label{fig:each-patch}
\end{figure}

We extracted the spectrum from different patches of GTIs of each \textit{NuSTAR} observation epoch. Since the GTI-patches have low exposure time, we grouped the spectrum with only 30 counts per bin. Patches with an exposure time less than 500 seconds are merged together before extracting the spectrum. 
We did a simultaneous joint fitting of all the GTI-patches under each epoch using the model M1. In the joint fitting, all parameters are tied between the patches except \textit{gabs strength}. The line-of-sight column density, $n_H$, is frozen at $0.45$ $\times$ 10$^{22}$ cm$^{-2}$ found from broadband \textit{NICER-NuSTAR} spectral modelling.
In Fig.~\ref{fig:each-patch}, we plot the simultaneous joint fitting of the spectra for Epoch 9, which has 7 patches of GTIs. The black, red, green, blue, cyan, magenta and yellow colours represent them in the ascending order of time.
 
\begin{figure}
	\centering
		\includegraphics[width=0.48\textwidth]{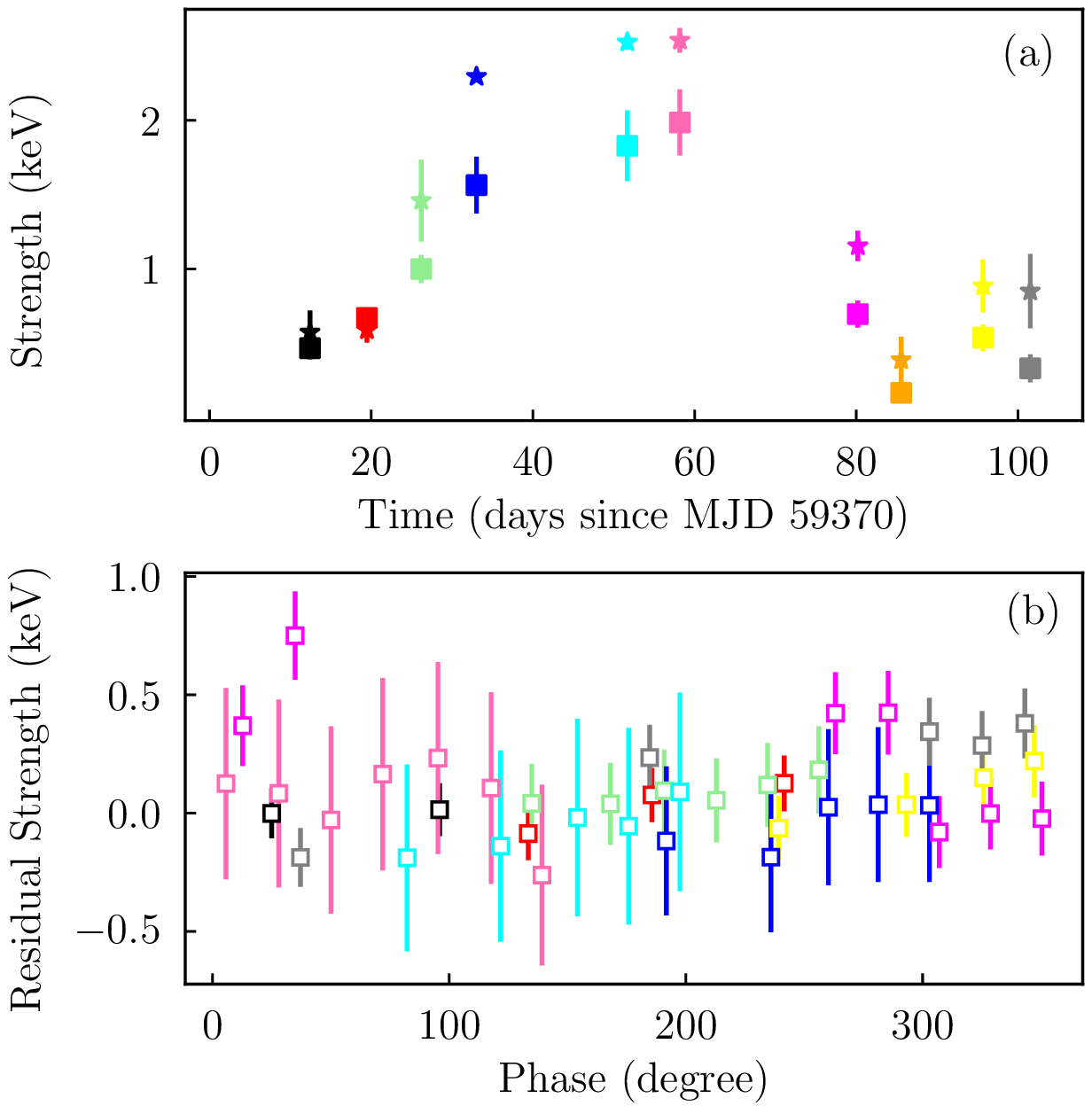} 
	\caption{\textbf{(a)} Evolution of \textit{gabs strength} (S$_i$) estimated from \textit{NuSTAR} using the models $tbabs(thcomp \times diskbb)edge \times gabs$ (square symbol) and $tbabs(diskbb+relxilllp)\times gabs$ (star symbol). 
	\textbf{(b)} Residual strength (S$_p$ - S$_i$) for different patches in each \textit{NuSTAR} epoch with the orbital phase. The colours black, red, green, blue, cyan, pink, magenta, orange, yellow and grey represent the \textit{NuSTAR} epochs in chronological order. See text for details.}
	\label{fig:bh-orbit}
\end{figure}

We also fitted all the 10 epochs (Table~\ref{tab:allobs}) of \textit{NuSTAR} observations using the models M1 and M2 discussed before. We calculated the \textit{gabs strength} (S$_i$) for each \textit{NuSTAR} epoch using both models. The evolution of S$_i$ is shown in Fig.~\ref{fig:bh-orbit}a. 
 The colours black, red, green, blue, cyan, pink, magenta, orange, yellow and grey indicate the NuSTAR epochs in chronological order. The \textit{gabs strength} of \textit{NuSTAR} data is showing the same trend of wideband spectral data; reaching maximum on Epoch 9 (pink in colour). The absorption \textit{strength} represented by the square symbol is estimated using the model M1, whereas the same using M2 are denoted by the star symbol. We notice that the value estimated using M2 are marginally higher than the same from M1. This can be the effect of an additional \textit{edge} component used in the M1 model.
%The value of different $S_p$s in each epoch is consistent with its $S_i$. 
%The values of $S_p$ are lying within the error ranges of each other for a same epoch. 
Similarly, we estimated the \textit{gabs strength} (S$_p$) corresponding to each GTI-patch for a given epoch from the patch-spectra modelling using M1. 
The residual \textit{strength} (S$_p$ - S$_i$) is measured for each patches inside an epoch and are plotted against the orbital phase in Fig.~\ref{fig:bh-orbit}b. 
The binary orbital period (P) of 4U $1543-47$ is  $26.79377\pm0.00007$ hours \citep{1998ApJ...499..375O,2003IAUS..212..365O}. The orbital position of each \textit{NuSTAR} patch has been identified based on the start time (MJD 59382.42) of Epoch 1 as the reference time.
For Epoch 12 (data in orange colour in Fig.~\ref{fig:bh-orbit}a), the value of S$_i$ is unusually low, and the estimated S$_p$ from the patch-spectra modelling of this epoch is not reliable. Therefore, we ignore Epoch 12 from Fig.~\ref{fig:bh-orbit}b.
The residual varies within $\pm 0.5$ keV (except one patch), and we see only a marginal variation (within uncertainties) in \textit{strength} within an orbit. This implies that the orbital position of the BH and the companion is not responsible for the dynamic nature of the absorption features. In fact, we do not expect such behaviour for a low inclination system.

The X-ray luminosity of the source at the peak (see Epoch 1 in Table~\ref{tab:pheno-ni-nu-param}) of the outburst is extremely high, and it may irradiate (see \cite{2001NewAR..45..449L} for a review) the outer accretion disk and the companion star. If the irradiation affects the companion star, either a fresh accretion of matter starts at the hot spot or enhances the stellar wind in the companion star. The former may produce multiple triggering in the same outburst event, which has been observed, for example, in GX $339-4$ \citep{2019MNRAS.486.2705A}. 
If the latter happens, the highly ionized wind material may absorb the X-ray emission from the primary to produce the broad absorption feature. The companion of 4U $1543-47$ is an A2V type star with a mass M$_c$=2.45 M$_\odot$ \citep{2006MNRAS.371.1334R} and radius R$_c$=2.84 R$_\odot$ \citep{1998ApJ...499..375O}. The escape velocity ($v_e$) of the stellar material from the surface of the companion is calculated as 573 km s$^{-1}$. The binary separation (a) between the BH and the companion is estimated as 7.18 $\times 10^{11}$ cm by considering a BH mass, $\rm M_{BH}$=9.4 M$_\odot$ \citep{2006MNRAS.371.1334R} using the relation $\rm P^2$/$\rm a^3=$ 4$\pi^2$/ $ G(\rm M_{BH}+ \rm M_c)$, where $G$ is the Gravitational constant. We observe that the stellar wind takes only a few hours to reach the primary. The column density of stellar wind and the ionization state should reduce with the decrease of the X-ray luminosity of the primary. Therefore, we expect that the strength of the broad absorption feature should reduce along with the progress of the outburst. Instead, we observe that the absorption strength enhances and becomes strongest during Epoch 9. Moreover, the estimated stellar wind speed is not sufficient
to blue shift the highest ionized lines of Fe XXVI, to produce the observed absorption feature. Therefore, the stellar wind has no role in the dynamic absorption features in the spectra of 4U $1543-47$.

The irradiation of the outer accretion disk enhances the accretion rate; therefore, the outburst source stays in the high luminosity state for a longer duration \citep{1998MNRAS.296L..45K,2001NewAR..45..449L,2019MNRAS.486.2705A,2020A&A...637A..47A}. This is possibly causing 2021 outburst of 4U $1543-47$ to decline very slowly (over $\sim$ 175 days, see Fig.~\ref{fig:lc-hr}). The super Eddington peak luminosity (Epoch 1 in Table~\ref{tab:pheno-ni-nu-param}) of the source can launch strong disk-wind (e.g., \citet{2015ApJ...813L..37K,2018MNRAS.479.3987M}). The presence of the accretion disk-wind is more prominent in the soft state of X-ray binaries \citep{2008ApJ...680.1359M,2009Natur.458..481N,2012MNRAS.422L..11P}, though disk-winds are not exclusively confined to soft spectral state \citep{2002ApJ...567.1102L}. 
Spectral analysis of 4U $1543-47$ 
(Table~\ref{tab:pheno-ni-nu-param}) suggests that the source was in the HSS during our study. Also, we notice spectral softening happens till Epoch 9 ($\sim$ day 60) and beyond which spectra 
gradually become harder (Table~\ref{tab:pheno-ni-nu-param} $\&$ Table~\ref{tab:refl-ni-nu-param}). If the disk-ionized winds are responsible for the absorption features, then the strength of the features would be maximum when the source is softer. Surprisingly the \textit{strength} of the absorption feature is maximum on $\sim$ day 60 as per our analysis (Fig.~\ref{fig:bh-orbit}a), and it is keeps-on decreasing further. The optical depth ($\tau_0$) evolution (Fig.~\ref{fig:plot-pheno}c) also suggests that the absorption column is maximally populated on $\sim$ day 60. 
 
The transition energy of the most ionized line with the highest absorption yield, i.e., Fe XXV and Fe XXVI, are 6.68 keV and 6.97 keV respectively (provided by \texttt{XSTAR} line finding list\footnote{\url{https://heasarc.gsfc.nasa.gov/docs/software/xstar/xstar.html}}). Assuming the absorption feature (with \textit{line E} $\sim$ 10 keV) in the \textit{NuSTAR} spectra is produced due to the absorption of the accretion disk photons by the highly ionized blue shifted disk-wind, the estimated wind speed is reaching 30\% of the speed of light to blue shift the Fe XXVI line energy to 10 keV. Such a fast disk-wind has never been observed in X-ray binary systems. In fact, highly ionized wind (say, Fe XXVI) is never detected  \citep{2012MNRAS.422L..11P} in BH-XRB systems with low inclination angle; for example, GX $339-4$, XTE J$1817-330$, 4U $1957+115$, XTE J$1650-500$, GRS $1758-258$ etc. Therefore, this detection is the first of its kind for X-ray binaries.
However, mildly relativistic disk-wind is not uncommon in quasars and AGNs \citep{2009ApJ...701..493R,2015Natur.519..436T,2017MNRAS.468.1442H}. The other difficulty is the width ($\sigma$) of the absorption feature, which is as broad as 2 keV on Epoch 9 (Table~\ref{tab:pheno-ni-nu-param}). Known line-broadening processes due to turbulence or scattering will face serious challenges in explaining the line width if it is from a single line. Instead, it is more likely that the broad feature can be produced by combining multiple lines of various ionization states of iron.

%We have seen that the phenomenological spectral fitting of \textit{NICER-NuSTAR} data with model M1 reveals that there exist a n

The phenomenological spectral fitting of \textit{NuSTAR} data with model M1 reveals the presence of neutral Fe K$-\alpha$ absorption edge and the broad ionized absorption features. %same as that observed in \S\ref{sec:pheno-modelling}.
We calculate the equivalent width (EW), which is a measure of the strength of an absorption line, of both absorption features to find if there exists any connection between these two components. The EW is defined as \citep{2018ApJ...869...97A},
\begin{equation}
    EW= \int_{0}^{\infty} [1- F(E) ] \; dE.
\end{equation}
 The energy dependent function $F(E)$ for the \textit{gabs} component is given by,
\begin{equation}
    F(E)= \rm exp(- \tau); \quad \tau= \tau_0 \; \rm exp \left[ - (E - E_g)^2/2 \sigma ^2 \right],
\end{equation}
where $\tau_0$, E$_g$ and $\sigma$ are the optical depth, line energy and line width respectively.
Similarly, $F(E)$ corresponds to the \textit{edge} component in XSPEC is given by,

\begin{equation}
   I(E)=
   \begin{cases} 
       1 & \text{if  E $\leq$ E$_e$} \\
       \rm exp[- D\, (E/E_e)^{-3} ]  & \text{if  E $\geq$ E$_e$,} 
   \end{cases}
\end{equation}
where E$_e$ and D are the threshold energy and absorption depth, respectively.

The evolution of EW calculated based on \textit{NuSTAR} phenomenological modelling is shown in Fig.~\ref{fig:ew} for both \textit{edge} (red in colour) and  \textit{gabs} (blue in colour) components.
The \textit{gabs} EW  increases till Epoch 9 and then gradually decline, except for Epoch 12 (Fig.~\ref{fig:bh-orbit}a) \& 13  (Table~\ref{tab:pheno-ni-nu-param}). The implication of this result and the connection between both components (\textit{gabs} and \textit{edge}) are discussed in \S\ref{sec:discuss}.

\section{Discussion}
\label{sec:discuss}
    The wideband spectral modelling of \textit{NICER-NuSTAR} and \textit{AstroSat} data reveals that the inner disc temperature $T_{in}$ is highest (1.27 keV) on Epoch 1, and it keeps on decreasing during the decay of the outburst (Fig.~\ref{fig:plot-pheno}a). The estimated \textit{diskbb norm} (Table~\ref{tab:pheno-ni-nu-param}, Table~\ref{tab:refl-ni-nu-param}) suggests a marginal inward movement of the inner disk radius $r_{in}$ since \textit{diskbb norm} $\propto$ $r_{in}^2$. However, the decrease in $r_{in}$ could not prevent the drop in $T_{in}$ due to the gradual decline in $\dot M$ as $T_{in}$ $\propto$ $\dot M^{1/4}$ $r_{in}^{-3/4}$ \citep{frank_king_raine_2002}. %\textcolor{magenta}{ Therefore, the change in T$_{in}$ mostly controlled by the accretion rate.} 
   The extreme luminosity in the inner disk may slow down the accretion of matter to the BH. On the other hand, the high accretion disk luminosity (Table~\ref{tab:pheno-ni-nu-param}) can irradiate the outer accretion disk, enhancing the accretion of matter. If most of the accreted matter is released as the disk-wind, the amount of matter actually transfers to the inner disk for falling onto the BH is much less. Therefore, the gradual decline of T$_{in}$ is due to the reduction of effective infall of matter onto the BH through the inner disk. 
     The source luminosity is completely soft-photons-dominated due to very little fractional Comptonization (\textit{cov$\_$frac} in Table~\ref{tab:pheno-ni-nu-param}), low corona temperature, and steeper photon index $\Gamma$ (Fig.~\ref{fig:plot-pheno}b). Therefore, the source was in the high/soft spectral state during our study, and it was the softest on Epoch 9.
    
The important parameters for reflection modelling are shown in Table~\ref{tab:refl-ni-nu-param} and in Fig.~\ref{fig:plot-refl}. The lamp-post height comes closer to the central object as the source becomes softer. The reflection fraction ($R_f$) increases and hits the boundary when the source is softest. 
If the value of ionization parameter, log $\xi$ $\gtrsim 3$, the fluorescence yield of the highly ionized Fe line (more ionized than Fe XXIII) increases \citep{1993MNRAS.262..179M}. The high value of log $\xi$ (Table~\ref{tab:refl-ni-nu-param}) obtained from reflection modelling suggests a highly ionized accretion disk throughout the outburst. Combining all the factors like extreme luminosity, high log $\xi$, and an overabundance of Fe (parameter $A_{Fe}$ in Table~\ref{tab:refl-ni-nu-param}) refer to a highly ionized disk-wind having a significant yield of Fe XXV, Fe XXVI etc.
\begin{figure}
	\centering
		\includegraphics[width=0.48\textwidth]{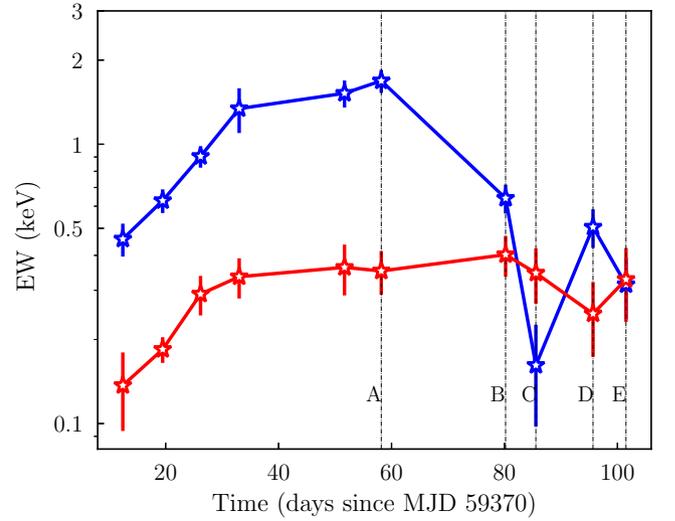} 
	\caption{Evolution of equivalent width of \textit{gabs} (blue in colour) and \textit{edge} (red in colour) components from  the \textit{NuSTAR} epochs. The vertical lines (A-E) are used to explain the figure in the description.}
	\label{fig:ew}
\end{figure}

An important characteristic of this source is the presence of a broad, symmetric and dynamic absorption feature in the spectrum $\sim$ $8-11$ keV. 
We presented various possibilities regarding the origin of this absorption feature in \S\ref{sec:absorption}. Finally, we concluded that the fast moving ionized disk-wind could absorb the primary X-ray photons and produce this broad absorption feature. The phenomenological modelling shows the presence of the neutral Fe K$\alpha$ absorption (\textit{edge}) at $\sim$ $7.1-7.4$ keV, originating from the outer part of the accretion disk. 
The initial steep rise of the EW of the neutral component (red star in Fig.~\ref{fig:ew}) indicates the enhancement of disk matter due to irradiation of the outer disk. Due to the availability of more matter, the radiation pressure of the highly luminous inner part could release more ionized matter.
The source enters to the HSS, and the disk-wind gradually becomes very active till Epoch 9 (possibly Epoch 10 also) where the \textit{gabs} EW is maximum (marked A in Fig.~\ref{fig:ew}) and the disk spectrum is the softest (Fig.~\ref{fig:plot-pheno}b).
Therefore, the evolution of the ionized EW (blue square in Fig.~\ref{fig:ew}) followed the neutral component till day 60 (Epoch 9). After that, EW of the ionized component declines gradually (AB in Fig.~\ref{fig:ew}) because the disk luminosity has reduced significantly, and a good fraction of inner disk matter has already been lost in the form of wind. %\textcolor{green}{\textit{NuSTAR} data on Epoch 12 reveals a sudden drop of the strength (orange points in Fig.~\ref{fig:bh-orbit}a) and EW (marked `C' in Fig.~\ref{fig:ew}) of \textit{gabs} component. }
The neutral component remains unaffected as it takes a viscous timescale to propagate the same to the outer disk. The \textit{NuSTAR} data on Epoch 12 reveals a sudden drop of the strength (orange points in Fig.~\ref{fig:bh-orbit}a) and EW (marked `C' in Fig.~\ref{fig:ew}) of \textit{gabs} component. We also notice the same signature in the \textit{AstroSat} data on Epoch 13, observed after 2 days of the \textit{NuSTAR}'s Epoch 12 observation; the \textit{gabs strength} (Table~\ref{tab:pheno-ni-nu-param}) on Epoch 13 is smaller by a few factor compared to the nearby observations. We identify this sudden drop of ionized EW can be due to the evacuation of the inner disk. The huge central luminosity may slow down the accretion onto the central object, and most of the accreted matter is released through disk-wind. Once the disk luminosity reduces, there is a sudden infall of matter onto the BH, leading to an evacuation of the inner disk.

If this interpretation is correct, we expect a relatively harder spectrum due to a significant drop of soft photon flux during Epochs 12 \& 13. To characterise this, we calculate the observed flux in $0.5-7$ keV (soft) and $7-20$ keV (hard) band for \textit{NuSTAR} and \textit{AstroSat} data. The \textit{AstroSat} soft and hard fluxes on Epoch 10 are $7.45 \times 10^{-8}$ erg cm$^{-2}$ s$^{-1}$ and $1.23 \times 10^{-9}$ erg cm$^{-2}$ s$^{-1}$ respectively. The drop of \textit{AstroSat} soft flux on the next observation (Epoch 13) is $14\%$, whereas the hard flux increases by $5\%$. This resulted in a relatively harder spectral index (marked by a blue circle in Fig.~\ref{fig:plot-pheno}b) on Epoch 13.
Similarly, the \textit{NuSTAR} soft and hard fluxes on Epoch 11 are $6.48 \times 10^{-8}$ erg cm$^{-2}$ s$^{-1}$ and $1.1 \times 10^{-9}$ erg cm$^{-2}$ s$^{-1}$ respectively. The drop of \textit{NuSTAR} soft flux on Epoch 12 is $10\%$, whereas the hard flux increases by a factor of 2. Therefore, the suddenly enhanced accretion at the inner disk resulted in these dramatic changes in the EW and spectral properties. However, the inner accretion disk recovers over the next 10 days (marked CD) due to the transfer of matter from the outer disk, and the ionized component returns back to a gradual declination. Interestingly, the neutral component (or the outer disk) follows the same trend as the ionized component, namely the decline and refilling signature (red stars between BE), with a delay of the typical viscous timescale of $10-15$ days.

\section{Summary and Conclusion}
\label{sec:concl}
We study the wideband spectral properties of the 2021 outburst of 4U $1543-47$. The \textit{MAXI/GSC} lightcurve (Fig.~\ref{fig:lc-hr}) shows that the outburst rises over 9 days followed by a slow decay over $\sim$ 175 days. %The source was extremely bright (super-Eddington peak luminosity) during the outburst. 
We use multi-instruments data (\textit{NICER, NuSTAR and AstroSat}) for simultaneous broadband spectral study over a period of 100 days from MJD 59370.
We have performed the spectral study using the phenomenological model M1 and reflection model M2. The major findings from our study are summarized below:

\begin{itemize}
    \item The source generally remains very bright during this outburst with a super Eddington peak luminosity on Epoch 1 (Table~\ref{tab:pheno-ni-nu-param}). 
    %Hence it can launch strong disk-wind.
        \item The source was in the HSS during our study, with a steep photon index (Fig.~\ref{fig:plot-pheno}b) due to a very small fraction ($< 3\%$) of inverse-Comptonized photons and low corona temperature.
        \item The reflection modelling reveals that the inclination of the system is between 32$^\circ - $40$^\circ$. 
        %The reflection modelling reveals a highly ionized disk (log $\xi$ > 3) and overabundance of iron ($3.6-10$ $A_{Fe,\odot}$) r study. 
        \item The extreme luminosity, high ionization (log $\xi$ > 3) and overabundance of iron ($3.6-10$ $A_{Fe,\odot}$) indicate the presence of disk-wind with a significant yield of highly ionized iron species.
        \item Presence of a broad, dynamic absorption feature at $\sim 8-11$ keV is observed throughout our study. This detection is the first of its kind for X-ray binaries. We propose that this  feature is due to the absorption of the accretion disk photons by the highly ionized, blue shifted disk-wind. The strength of the ionized absorption feature (Table~\ref{tab:pheno-ni-nu-param} \& Table~\ref{tab:refl-ni-nu-param}) increases between Epoch 1 to Epoch 9 as the disk-wind column density is expected to increase with the spectral softening of the source. %and decreases gradually (except Epoch 12 $\&$ 13) afterwards.
%\textcolor{green}{and we observed an increase in the strength of the absorption feature (Table~\ref{tab:pheno-ni-nu-param} \& Table~\ref{tab:refl-ni-nu-param}) between Epoch 1 to Epoch 9 which is a consequence of it.}
        
   \item The observed line energy of the absorption feature suggests an estimated wind speed of nearly 30$\%$ of the speed of light to blue shift the most ionized line with the highest absorption yield like Fe XXVI. Hence it would become the first X-ray binary source to show a highly relativistic disk-wind.
    \item The initial steep rise of the neutral component EW (red star in Fig.~\ref{fig:ew}) is an indication of the enhancement of disk matter due to irradiation of the outer disk. It enhances the accretion rate and hence the source remains in the high luminosity state and decays very slowly.
    \item The evolution of EW (Fig.~\ref{fig:ew}) of the neutral absorption component (\textit{edge}) and the same of the ionized component (\textit{gabs}), follow each other with a delay of the typical viscous timescale of $10-15$ days.
    %    \item All these results infer that the broad absorption feature originated due to the absorption of disk photons by the fast-moving ionized disk-wind.
        \item An evacuation of the inner accretion disk is observed during Epoch $12-13$. This event leaves a signature of the drop in the soft photon flux and an enhancement of hard flux. Therefore, the spectrum becomes relatively harder (blue circle in Fig.~\ref{fig:plot-pheno}b).
\end{itemize}

Finally, this study suggests that accretion dynamics of 4U $1543-47$ during 2021 outburst is regulated by the disk-wind.

\section*{Acknowledgements}
The authors wish to thank the anonymous reviewer for the insightful suggestions which significantly improved the quality of the publication.
This work uses data from the \textit{NICER} and \textit{NuSTAR} mission by the National Aeronautics and Space Administration. This work also has used data from the \textit{AstroSat} mission of the ISRO archived at the Indian Space Science Data Centre (ISSDC). The work has been performed utilizing the calibration databases, and auxiliary analysis tools developed, maintained and distributed by \textit{AstroSat-SXT} team with members from various institutions in India and abroad. The High Energy Astrophysics Science Archive Research Center (HEASARC), which provides the software and NASA’s Astrophysics Data System Bibliographic Services are also acknowledged. BGR acknowledges the financial support of ISRO under AstroSat archival data utilization program Sanction order No. DS-2B-13013(2)/13/2019-Sec.2. AN thanks GH, SAG, DD, PDMSA, and Director, URSC for the support to carry
out this research.

%%%%%%%%%%%%%%%%%%%%%%%%%%%%%%%%%%%%%%%%%%%%%%%%%%
\section*{Data Availability}

The data from \textit{NICER} and \textit{NuSTAR} underlying this article are available in HEASARC, at \url{https://heasarc.gsfc.nasa.gov/docs/archive.html}.
\textit{AstroSat} data archive is available at \url{https://astrobrowse.issdc.gov.in/astro_archive/archive/Home.jsp}.

%%%%%%%%%%%%%%%%%%%%%%%%%%%%%%%%%%%%%%%%%%%%%%%%%%

%%%%%%%%%%%%%%%%%%%% REFERENCES %%%%%%%%%%%%%%%%%%

\bibliographystyle{mnras}
\bibliography{4u} % bibtex is geethu.bib

%%%%%%%%%%%%%%%%% APPENDICES %%%%%%%%%%%%%%%%%%%%%

%%%%%%%%%%%%%%%%%%%%%%%%%%%%%%%%%%%%%%%%%%%%%%%%%%

% Don't change these lines
\bsp	% typesetting comment
\label{lastpage}
\end{document}